\documentclass[prl, aps, amsmath, amssymb, amsfonts,twocolumn]{revtex4}
\usepackage[german,american,english]{babel}
\usepackage{graphicx}
\usepackage{graphics}
\usepackage{dcolumn}
\usepackage{bm}
\usepackage{amssymb}
\usepackage{amsmath}
\usepackage{amsfonts}
\usepackage{epsfig}
\usepackage{mathbbol}
\usepackage{dcolumn}
\usepackage{subfigure}
\usepackage{hyperref}
\usepackage{float}
\hypersetup{                    
    colorlinks,
    citecolor=blue,
    filecolor=blue,
    linkcolor=blue,
    urlcolor=blue
}

 \newcommand {\beq}{\begin{equation}}
\newcommand {\eeq}{\end{equation}}
\newcommand {\beqn}{\begin{eqnarray}}
\newcommand {\eeqn}{\end{eqnarray}}
\newcommand {\bit}{\begin{itemize}}
\newcommand {\eit}{\end{itemize}}

\newcommand{\ba}{\begin{array}{rl}}
\newcommand{\ea}{\end{array}}
\newcommand{\bc}{\begin{cases}}
\newcommand{\ec}{\end{cases}}

\newcommand{\dps}{\displaystyle}

\newcommand{\om}{\iffalse}

\begin{document}
\title{Anomalous Hall Coulomb drag of massive Dirac fermions}

\author{Hong Liu, Weizhe Edward Liu, Dimitrie Culcer}
\affiliation{School of Physics, The University of New South Wales, Sydney 2052, Australia}
\begin{abstract}
Dirac fermions are actively investigated, and the discovery of the quantized anomalous Hall effect of massive Dirac fermions has spurred the promise of low-energy electronics. Some materials hosting Dirac fermions are natural platforms for interlayer coherence effects such as Coulomb drag and exciton condensation. Here we determine the role played by the anomalous Hall effect in Coulomb drag in massive Dirac fermion systems. We focus on topological insulator films with out-of plane magnetizations in both the active and passive layers. The transverse response of the active layer is dominated by a topological term arising from the Berry curvature. We show that the topological mechanism does not contribute to Coulomb drag, yet the longitudinal drag force in the passive layer gives rise to a transverse drag current. This anomalous Hall drag current is independent of the active-layer magnetization, a fact that can be verified experimentally. It depends non-monotonically on the passive-layer magnetization, exhibiting a peak that becomes more pronounced at low densities. These findings should stimulate new experiments and quantitative studies of anomalous Hall drag. 
\end{abstract}
\date{\today}
\maketitle

The past decade has witnessed an energetic exploration of Dirac fermions in materials ranging from graphene \cite{KaneMele_QSHE_PRL05} to topological insulators \cite{Hasan_TI_RMP10}, transition metal dichalcogenides \cite{Dixiao_MoS2_2012} and Weyl and Dirac semimetals \cite{Burkov_Weyl_TI,Fuhrer_Na3Bi_2016,Fuhrer_NL_Dirac_semi}. Dirac fermions in 2D are described by the Hamiltonian $H_{\text{D}} = A \,{\bm \sigma}\cdot({\bm k}\times\hat{{\bm z}}) + M\sigma_z$, with ${\bm \sigma}=(\sigma_x,\sigma_y,\sigma_z)$ the usual Pauli matrices, ${\bm k} = (k_x, k_y)$ the 2D wave vector, $A$ stems from the Fermi velocity and $M$ a generic mass term. In the limit $M\rightarrow0$ the quasi-particle dispersion is linear, a feature that has aroused intense interest experimentally \cite{Wang_Bi2Te3_Ctrl_AM11, Benjamin_nature_2011,He_Bi2Te3_Film_WAL_ImpEff_PRL11,Ren-PRB-2012, Kim_sur_2012, Jinsong-Zhang-Yayu-Wang-2013-science, Lucas_nl_tran_2014, Jianshi_nl_spin_polarized_2014, Kozlov_prl_2014, Fuhrer_nature_2013, Cacho_prl_2015, ZhangJinsong_prb_2015, Hellerstedt_APL_2014,Kastl_surface_tran_nature_2015, E_control_SOT_TI} and theoretically \cite{Hwang_Gfn_Screening_PRB07,JungMacDonald_graphene_PRB2011,Durst_2015,Haizhou_WL_2014,FP_Bi2X_3_NewJ, Shi_Rappe_nl_2016, Culcer_TI_AHE_PRB11,Adam_2D_Tran_prb_2012,Yoshida-PRB-2012,LiQiuzi_2013_prb,Weizhe_TITF_2014_prb,Hai-Zhou_Conductivity_2014_prl,Das_prb_2015}. These studies have illuminated the considerable potential of Dirac fermions for spintronics, thermoelectricity, magnetoelectronics and topological quantum computing \cite{Das_Sarma_RMP_TQC}. 

Certain materials hosting Dirac fermions, such as 3D topological insulator (TI) slabs, are inherently two-layer systems naturally exhibiting interlayer coherence effects such as Coulomb drag \cite{1977_drag, Zheng-2DES-1993, Drag_riview}, in which the charge current in one layer \textit{drags} a charge current in the adjacent layer through the interlayer electron-electron interactions. Drag geometries are intensively studied experimentally in Dirac fermion systems as part of the search for exciton condensation \cite{Drag-dot, 1D-1D_science_Luttinger,1D-1D_prb_Dmitriev, edge_drag_prb, ee_drag_93_prb, Nandi_eh_nat, E-H_exciton_2016, Disorder_drag, Drag_semi_1996,Semi_drag_prl, 1995_Hall_drag,drag_metal_prb, Tse2007, Amorim_drag, Kim2011,Tse_SO_Drag_PRB07, Katsnelson2011, Hwang2011, M.Carrega2012, Narozhny_drag_2012, Kim2012, Gorbachevi_nature_2012, Sch_drag_prl_2013, Gamucci_nature_2014, Hall_drag_Graphene, Song_Halldrag_2013, Scharf_Alex_drag_prb_2012, Drag_graphnen_nl, Drag_TI_2013_Efimkin, Hong_drag_2015}. The most promising Dirac fermion materials have been magnetic TI slabs, in which a dissipationless quantized anomalous Hall effect has been discovered  \cite{Murakami_SHI_PRL04, QSH-HgTe_2007, Chang_QAHE_exper_Science2013, Precise-QAHE-CZChang}, which has already been harnessed successfully \cite{D.Reilly_USYD}, stimulating an intense search for device applications. The time-reversal symmetry breaking required in Hall effects \cite{Nagaosa-AHE-2010, Culcer_TI_AHE_PRB11, Valley_prb_Niu,Yu_TI_QuantAHE_Science10,Dixiao_MoS2_2012,Bi-graphene_massive_prl} gives Dirac fermions a finite mass and results in a non-trivial Berry curvature \cite{Dixiao_MoS2_2012,LuZhao_TITF_transport_PRL2013, Valley-Hall_graphene_prl_2007}. Coulomb drag of massive Dirac fermions is thus directly relevant to ongoing experiments, and raises important questions: if topological terms are present in the drag current they could be exploited in longitudinal transport, enabling a topological transistor. 

\begin{figure}
\begin{center}
\includegraphics[width=0.7\columnwidth]{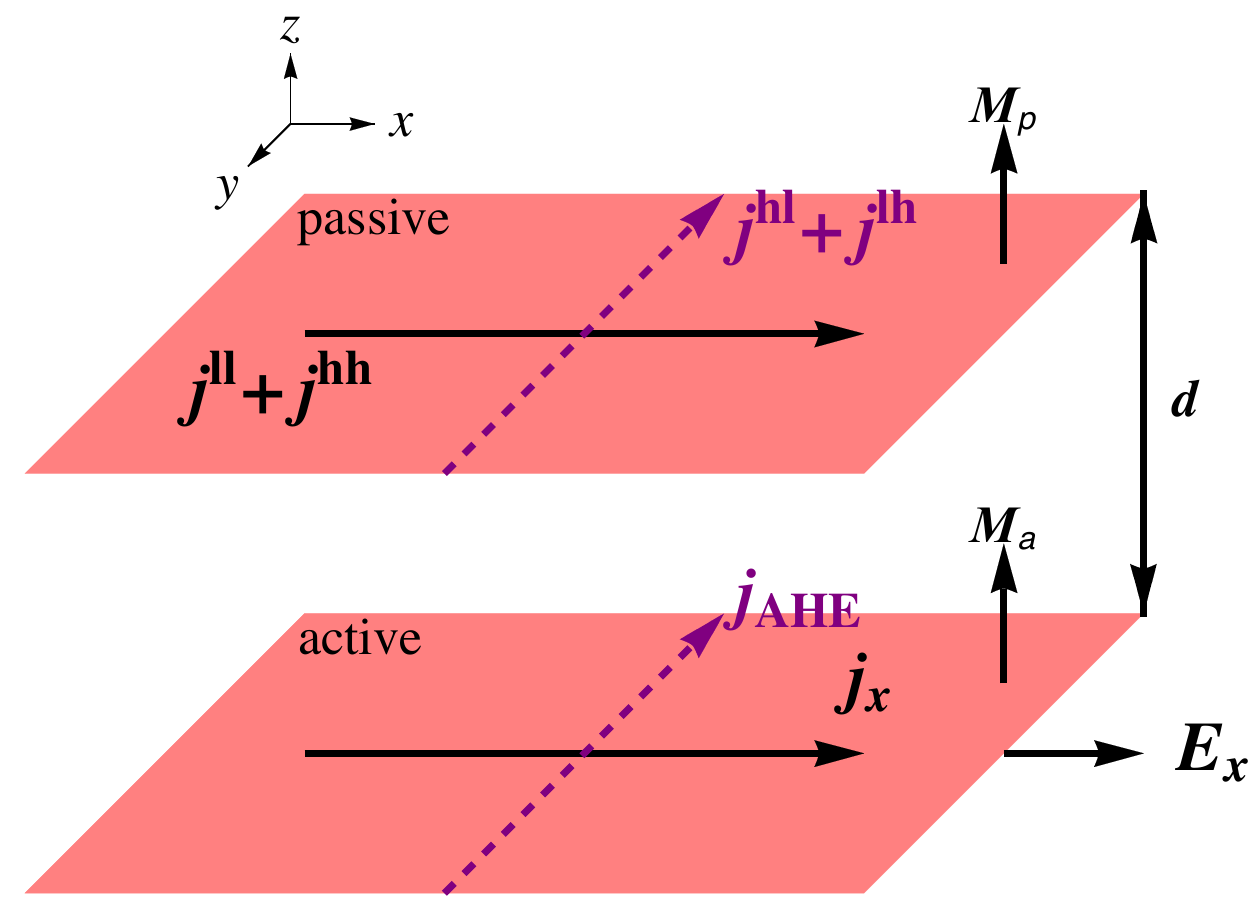}\\
\caption{\label{4-currents} Contributions to the drag current in magnetic TIs. The electric field gives rise to a longitudinal (${\bm j}_x$) and an anomalous Hall (${\bm j}_{\text{AHE}}$) current in the active layer. In the passive layer there are four contributions: ${\bm j}^{\text{ll}}$ is the longitudinal current dragged by ${\bm j}_x$; ${\bm j}^{\text{lh}}$ is the transverse current dragged by ${\bm j}_{\text{AHE}}$; ${\bm j}^{\text{hl}}$ is the anomalous Hall current generated by ${\bm j}^{\text{ll}}$; while ${\bm j}^{\text{hh}}$ is the anomalous Hall current generated by ${\bm j}^{\text{lh}}$, and it flows longitudinally. ${\bm M}_\text{a}$ and ${\bm M}_\text{p}$ are the magnetizations of the active and passive layers, respectively. $d$ is the layer separation.}
\end{center}
\end{figure}

Here we present a complete theory of Coulomb drag of massive Dirac fermions. The central findings of our work are that (i) the anomalous Hall current in the active layer does not generate a drag current in the passive layer (Fig.~\ref{4-currents}), and (ii) the remaining drag current only depends on $M_\text{p}$, the magnetization of the passive layer. Its dependence on $M_\text{p}$ is non-monotonic, with a peak at an intermediate value of $M_\text{p}$, which becomes pronounced at low densities. Although derived here using a minimal model for Dirac fermions, these results apply generally to materials with Rashba spin-orbit interactions. They stand in sharp contrast to conventional Coulomb drag in ordinary Hall systems \cite{1995_Hall_drag,1999_Hall_drag, Hall_drag_Graphene, Song_Halldrag_2013, Sch_drag_prl_2013}. There the Hall current in the active layer, caused by the Lorentz force rather than topology, makes a significant contribution to the drag response, which depends on the applied magnetic field. This work is intended to stimulate further experiments and quantitative studies on state-of-the-art samples, in which the above results can be verified.

We consider a TI film with both the top (active) and bottom (passive) surfaces magnetized either through doping with magnetic impurities or proximity coupling to ferromagnets. The magnetizations of the two surfaces are allowed to differ. Without loss of generality we assume (i) the carrier number density and hence the Fermi energy is the same in each layer and (ii) sidewall states do not participate in transport, an assumption that recent work has shown to be justified \cite{Tilahun_TIF_QHS_PRL2011}. The chemical potential lies in the surface conduction band of each layer. We require $\varepsilon_\text{F}\tau_l/\hbar\gg1$ in each layer $l \in \{\text{a} \equiv \text{active},\text{p} \equiv \text{passive}\}$, with $\varepsilon_\text{F}$ the Fermi energy located in the surface conduction bands and $\tau_l$ the momentum scattering time. The two-layer effective band Hamiltonian
\begin{equation}\label{band-H}
H_{0{\bm k}}=\tau_z\otimes h_{{\bm k}}+\text{diag}\{M_{\text{a}},-M_{\text{a}},M_{\text{p}},-M_{\text{p}}\},
\end{equation}
where the (Rashba) Hamiltonian of a single layer $h_{{\bm k}}= A\,{\bm \sigma}\cdot({\bm k}\times\hat{{\bm z}})\equiv -Ak{\bm \sigma}\cdot\hat{\bm \theta}$ with $\hat{\bm \theta}$ the tangential unit vector corresponding to ${\bm k}$. The Pauli matrix $\tau_z$ represents the layer degree of freedom. The eigenvalues of  Eq.~(\ref{band-H}) are $\varepsilon_{l\pm}={\pm}\sqrt{A^2k^2+M^2_{l}} \equiv \hbar\Omega_k^{(l)}$, the band index $s_{\bm k}=\pm$ with $+$ the conduction band and $-$ the valence band. 

We directly calculate the anomalous Hall drag resistivity $\rho^{yx}_\text{D}$: the procedure will be outlined below. We find ${\bm j}^{\text{lh}}$ and ${\bm j}^{\text{hh}}$ (Fig.~\ref{4-currents}) vanish identically, implying that the anomalous Hall current in the active layer does not generate a corresponding drag current in the passive layer. The physics can be understood in two steps. Firstly, an external electric field drives the electrons in the active layer longitudinally, and these in turn exert a longitudinal drag force on the passive-layer electrons. The drag force acts as an effective longitudinal driving term for the passive-layer electrons. The anomalous Hall component of the drag current, which at low temperatures can be sizeable compared to the longitudinal component, represents the transverse response of the passive layer to this effective longitudinal driving force, and depends on the passive-layer magnetization $M_\text{p}$. The electric field also generates an anomalous Hall current in the active layer. This response is dominated by topological terms of the order of the conductivity quantum, which represents a re-arrangement of charge carriers among spin-momentum locked energy states. Physically this is because Coulomb drag occurs as a result of the interaction between the charge densities in different layers, whereas the anomalous Hall current flowing in the active layer is not associated with a change in the charge density: it does not arise from a shift in the Fermi surface, but from the Berry phase acquired through the rearrangement of carriers among spin-momentum locked states. This implies that the anomalous Hall drag current is quite generally independent of the active-layer magnetization $M_\text{a}$. This can be easily verified experimentally. The corollary is that in a sample in which the active layer is undoped, so that it produces no longitudinal current but only a quantized anomalous Hall current, there will be no drag current at all in the passive layer.

We analyse the parameter dependence of $\rho^{yx}_\text{D}$ in Fig.~\ref{yxM} and Fig.~\ref{yx}. The relationship between $\rho^{yx}_\text{D}$ and the magnetization of the passive layer is illustrated in Fig.~\ref{yxM}. There is an upward trend at small magnetizations followed by a relatively slow decrease at larger values of $M_{\text{p}}$. The trend can be understood as follows. For $M_{\text{p}}\ll Ak_{\text{F}}$ one may expand Eq.~(\ref{jhl-c}) in $M_{\text{p}}$, which reveals that the current increases nearly linearly with $M_{\text{p}}$. In the opposite limit in which $M_{\text{p}}\gg Ak_{\text{F}}$ \cite{AHE_vertex_PRL_2006}, the anomalous Hall current vanishes, since the effect of spin-orbit coupling (chirality), which scales with $k_{\text{F}}$, becomes negligible. In fact at large $M_{\text{p}}$ one may expand the band energies in $Ak$, with the leading $k$-dependent term scaling as $k^2$, which suggests the system in this limit behaves as a regular, non-magnetic 2DEG. This explains the slow downward trend with increasing $M_{\text{p}}$, and hence the presence of the peak as a function of $M_{\text{p}}$. At larger electron densities the peak occurs at stronger values of the magnetization, since higher electron densities imply higher values of $Ak_\text{F}$, increasing the effect of chirality at the expense of the magnetization. We note that magnetizations of magnetic TIs have been measured by superconducting quantum interference device (SQUID) magnetometers \cite{Hor_DopedTI_FM_PRB10,Jinsong-Zhang-Yayu-Wang-2013-science}. Because the momentum scattering time is in principle known from the longitudinal conductivity of each layer \cite{Cacho_TI_tau,Peng_natcom_2016_tau}, the trend exhibited by $\rho^{yx}_\text{D}$ as a function of the magnetization $M_{\text{p}}$ can be verified experimentally.

\begin{figure}\label{yxM}
\begin{center}
\includegraphics[width=7cm]{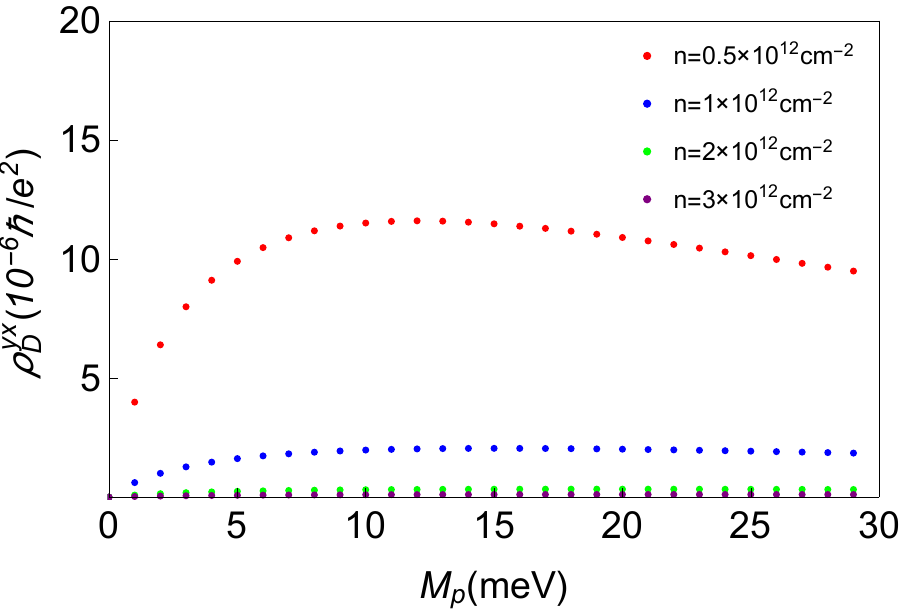}
\caption{\label{yxM} Magnetization dependence of
$\rho^{yx}_\mathrm{D}$ with $T=5~\text{K}$, $d=10~\text{nm}$, dielectric constant $\epsilon_r=20$, $A=4.1~\text{eV\AA}$ and transport time $\tau\approx0.1~\mathrm{ps}$.}
\end{center}
\end{figure}

We examine the dependence of the anomalous Hall drag resistivity on additional experimentally measurable parameters. In Fig.~\ref{yx}(a) the electron density dependence of $\rho^{yx}_{\text{D}}$ for interlayer separations $d=10,20,40 ~\text{nm}$ is shown. Compared with the longitudinal drag resistivity, the anomalous Hall drag resistivity has a weaker dependence on electron density. As compared with the longitudinal drag resistivity, the group velocity appearing in the susceptibility of the passive layer is replaced by the Berry curvature, leading to a weaker density dependence, yet no topological contribution. Next, Fig.~\ref{yx}(b) illustrates the temperature dependence of $\rho^{yx}_{\text{D}}$ for separations $d=10,20,40 ~\text{nm}$. We find that, much like $\rho^{xx}_{\text{D}}$, $\rho^{yx}_{\text{D}}$ also increases nearly quadratically with temperature. The $T^2$ dependence stems from the allowed phase space for electron-electron scattering at low temperature, and is expected for any interaction strength between the top and bottom layers of TIs, provided the carriers can be described using a Fermi liquid picture. Moreover, due to the absence of backscattering in TIs there is no correction logarithmic in temperature. Fig.~\ref{yx}(c) presents the layer separation dependence of $\rho^{yx}_{\text{D}}$ for three different Fermi wave vectors  $k_\text{F}= 0.4, 0.5, 0.6 \text{nm}^{-1}$. 

\begin{figure}\label{yx}
\begin{center}
\includegraphics[width=5.5cm]{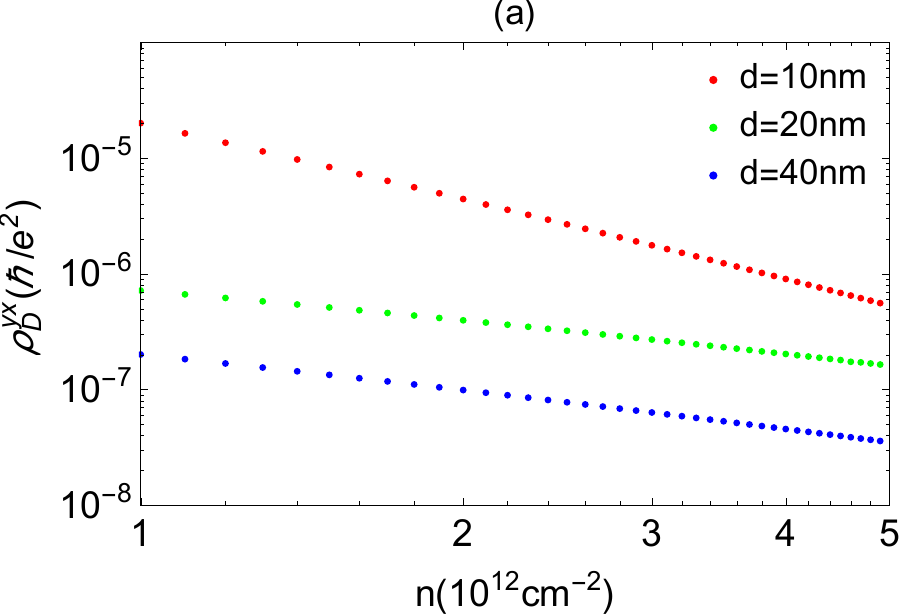}
\includegraphics[width=5.5cm]{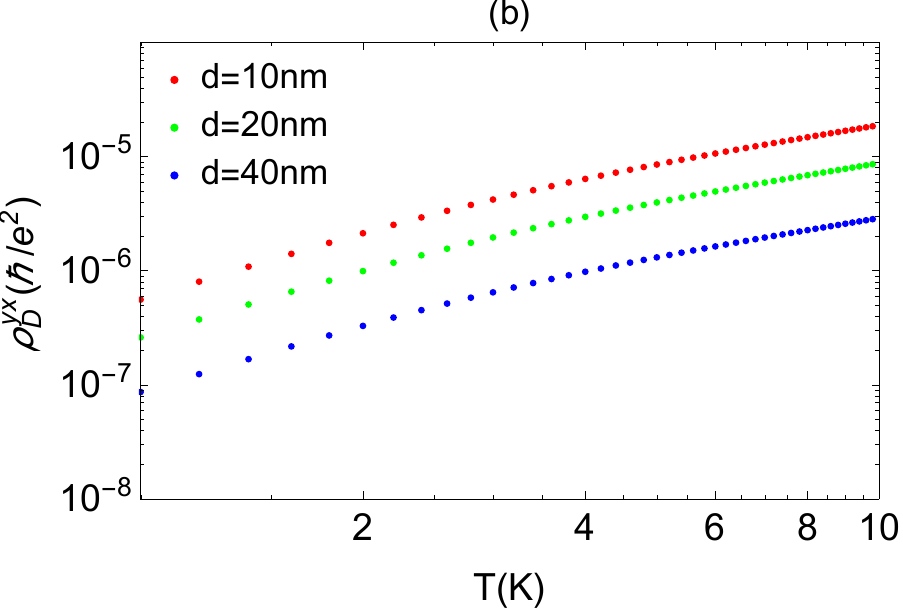}
\includegraphics[width=5.5cm]{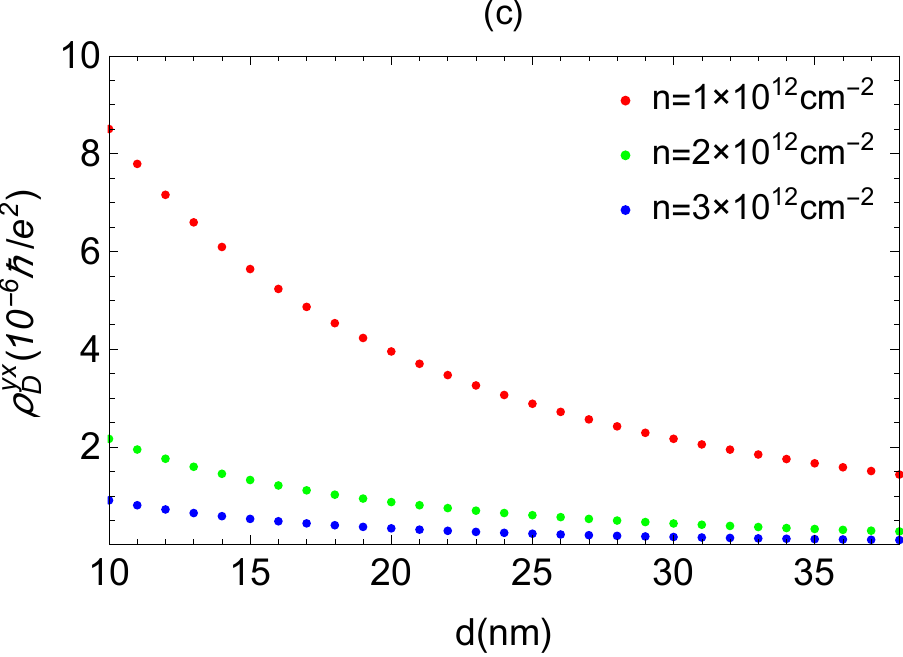}
\caption{\label{yx} (a) Electron density dependence at $T=5~\mathrm{K}$, (b) Temperature dependence with $n=10^{12}\text{cm}^{-2}$, (c) Layer separation dependence at $T=5~\mathrm{K}$. $M_{\text{p}} =10~\mathrm{meV}$, dielectric constant $\epsilon_r=20$, $A=4.1~\text{eV\AA}$ and transport time $\tau\approx0.1~\mathrm{ps}$.}
\end{center}
\end{figure}

We briefly summarize our formalism. The two-layer system is described by the many-body density matrix $\hat{F}$, which obeys the quantum Liouville equation \cite{Vasko}
\begin{equation}
\frac{\text{d}\hat{F}}{\text{d}t}+\frac{i}{\hbar}[\hat{H},\hat{F}]=0,
\end{equation}
where $\hat{H}=\hat{H}^{1e}+\hat{V}^{ee}$ with
$\hat{H}^{1e}\!=\!\sum_{\alpha\beta}H_{\alpha\beta}c^{\dag}_{\alpha}c_{\beta}$ and $\hat{V}^{ee}\!=\!\frac{1}{2}\sum_{\alpha\beta\gamma\delta}V^{ee}_{\alpha\beta\gamma\delta}c^\dag_\alpha c^\dag_\beta c_\gamma c_\delta$. The indices $\alpha\equiv{\bm k}s_{\bm k}l$ represent wave vector, band, and layer indices respectively. The matrix element $V^{ee}_{\alpha\beta\gamma\delta}$ in a generic basis $\{\phi_{\alpha}({\bm r})\}$ is given by
$V^{ee}_{\alpha\beta\gamma\delta}=\int \!d{\bm r}\int \!d{\bm r}' \ \phi^*_{\alpha}({\bm r})\phi^*_{\beta}({\bm r}')V^{ee}_{{\bm r}-{\bm r}'}\phi_{\delta}({\bm r})\phi_{\gamma}({\bm r}')$,
where $V^{ee}_{{\bm r}-{\bm r}'}$ is the Coulomb interaction in real space.  The one-particle reduced density matrix is the trace
$\rho_{\xi\eta}\!=\!\text{tr}(c^\dag_\eta c_\xi \hat{F})\equiv\langle c^\dag_\eta c_\xi\rangle\equiv\langle \hat{F}\rangle_{1e}$,
 which satisfies \cite{Culcer_TI_Int_PRB11}
 \begin{equation}\label{kinetic_original_0}
 \frac{\text{d}\rho_{\xi\eta}}{\text{d}t}+\frac{i}{\hbar}[\hat{H}_{1e},\hat{\rho}]_{\xi\eta}=\frac{i}{\hbar}\langle[\hat{V}_{ee},c^\dag_\eta c_\xi]\rangle,
 \end{equation}
 with averages such as $\langle[\hat{V}_{ee},c^\dag_\eta c_\xi]\rangle$ factorized as \cite{Vasko}
\begin{equation}\label{correlation}
\langle c^\dag_\alpha c^\dag_\beta c_\gamma c_\delta\rangle=\langle c^\dag_\alpha c_\delta\rangle\langle c^\dag_\beta c_\gamma\rangle -\langle c^\dag_\alpha c_\gamma\rangle\langle c^\dag_\beta c_\delta\rangle+ G_{\alpha\beta\gamma\delta}.
\end{equation}
The term $\hat{J}_{ee}(\hat{\rho}|t)$ in Eq.~(\ref{kinetic_original_0}) represents intralayer and interlayer electron-electron scattering. Since the intralayer electron-electron scattering does not contribute to the drag current, we concentrate on the interlayer electron-electron scattering,
\begin{equation}\label{electron-electron-m}
\ba
&\dps J^{\text{Inter,\text{e-e}}}_{{\bm k},\text{p},s_{\bm k}s'_{\bm k}} =\frac{\pi}{\hbar L^4}\sum_{{\bm k}_1{\bm k}'{{\bm k}'_1}}|v^{(\text{pa})}_{|{\bm k}-{\bm k}_1|}|^2\delta_{{\bm k}+{\bm k}',{\bm k}_1+{\bm k}'_1}\big\{\\[3ex]
&\dps \sum_{i=1}^{4}\!P^{(i)}_{s_{\bm k}s_{{\bm k}_1}s'_{\bm k}}A^{(i)}_{s_{\bm k'}s_{{\bm k}'_1}}\delta[\varepsilon^{(\text{p})}_{k_1,s_{{\bm k}_1}}\!-\varepsilon^{(\text{p})}_{k,s'_{\bm k}}\!+\varepsilon^{(\text{a})}_{k'_1,s_{{\bm k}'_1}}\!-\varepsilon^{(\text{a})}_{k',s_{\bm k'}}]+\\[3ex]
&\dps \sum_{i=5}^8\!P^{(i)}_{s_{\bm k}s_{{\bm k}_1}s'_{\bm k}}A^{(i)}_{s_{\bm k'}s_{{\bm k}'_1}}\delta[\varepsilon^{(\text{p})}_{k_1,s_{{\bm k}_1}}\!-\varepsilon^{(\text{p})}_{k,s_{\bm k}}\!+\varepsilon^{(\text{a})}_{k'_1,s_{{\bm k}'_1}}\!-\varepsilon^{(\text{a})}_{k',s_{\bm k'}}]\big\},
\ea
\end{equation}
where $L^2$ is the area of the 2D system. The interlayer momentum transfer ${\bm q}={\bm k}-{\bm k}_1={\bm k}'_1-{\bm k}'$, and $v^{(\text{pa})}_{|{\bm k}-{\bm k}_1|}$ is the interlayer Coulomb interaction. $A^{(i)}_{s_{\bm k'}s_{{\bm k}'_1}}$ and $P^{(i)}_{s_{\bm k}s_{{\bm k}_1}s'_{\bm k}}$ are given explicitly in Table~I. of the Supplement. 

The single-particle Hamiltonian $\hat{H}^{1e}=\hat{H}_0+\hat{H}_{E}+\hat{U}$, where $\hat{H}_0$ is the band Hamiltonian defined in Eq.~(\ref{band-H}), $\hat{H}_E=e(\hat{\bm E}\otimes\openone)\cdot\hat{\bm r}$ is the electrostatic potential due to the driving electric field with $\hat{\bm r}$ the position operator and $\hat{U}$ the disorder potential, which is assumed to be a scalar in spin space. The kinetic equation of the two-layer system
\begin{subequations}\label{eqab}
 \begin{equation}\label{active}\vspace{-0.5cm}
  \frac{\text{d}f^{(\text{a})}_{\bm k}}{\text{d}t}+\frac{i}{\hbar}[H^{(\text{a})}_{0{\bm k}},f^{(\text{a})}_{\bm k}]+\hat{J}_{0}(f^{(\text{a})}_{\bm k})=-\frac{i}{\hbar}[H^E_{{\bm k}},f^{(\text{a})}_{0{\bm k}}],
 \end{equation}
\begin{equation}\label{passive}
\frac{\text{d}f^{(\text{p})}_{\bm k}}{\text{d}t}+\frac{i}{\hbar}[H^{(\text{p})}_{0{\bm k}},f^{(\text{p})}_{\bm k}]+\hat{J}_{0}(f^{(\text{p})}_{\bm k})=J^{\text{Inter},\text{e-e}}_{{\bm k},\text{p}},
\end{equation}
\end{subequations}
with $H^{(l)}_{0{\bm k}}=h^{(l)}_{\bm k}+M_l \sigma_z$ the band Hamiltonian and $f^{(l)}_{0{\bm k}}$ the equilibrium density matrix of each magnetic layer. We neglect the interlayer electron-electron collision integral $J^{\text{Inter},\text{e-e}}_{{\bm k},\text{a}}$ in the active layer, and the electron-impurity scattering in the first Born approximation yields \cite{Culcer_TI_AHE_PRB11}
\begin{equation}
\hat{J}_{0}(f^{(l)}_{\bm k})=\bigg\langle\int^\infty_0\frac{\text{d}t'}{\hbar^2}[\hat{U},e^{-i\hat{H}t'/\hbar}[\hat{U},\hat{f}]
e^{i\hat{H}t'/\hbar}]\bigg\rangle_{{\bm k}{\bm k}},
\end{equation}
where $\langle\rangle$ denotes the average over impurity configurations. We write $f^{(l)}_{\bm k}=n^{(l)}_{\bm k}\mathbb{1}+S^{(l)}_{\bm k}$, with $S^{(l)}_{\bm k}$ a $2\times2$ Hermitian matrix which can be written in terms of the Pauli spin matrices. As in Ref.~[\onlinecite{Culcer_TI_AHE_PRB11}] we choose the unit vectors
$\hat{{\bm \Omega}}_{\bm k}=-a_k\hat{{\bm \theta}} + b_k\hat{{\bm z}}$,
$\hat{{\bm k}}_{eff}=\hat{{\bm k}}$ and 
$\hat{{\bm z}}_{eff}=a_k\hat{{\bm z}}+b_k\hat{{\bm \theta}}$,
with $a_k=2Ak/\hbar\Omega_k$, $b_k=2M_l/\hbar\Omega_k$ so that $a^2_k+b^2_k=1$ (we suppress the layer indices for simplicity) \footnote{Although magnetic impurities also cause spin-dependent scattering, our recent work showed that the anomalous Hall effect in TIs is dominated by the band structure spin-orbit coupling \cite{Culcer_TI_AHE_PRB11}, hence we do not include explicitly spin-dependent scattering due to the magnetic impurities.}. We project ${\bm \sigma}$ as $\sigma_{{\bm k},\parallel}={\bm \sigma}\cdot\hat{\bm \Omega}_{\bm k}$, $\sigma_{{\bm k},k}={\bm \sigma}\cdot\hat{\bm k}$ and $\sigma_{{\bm k},z_{\text{eff}}}={\bm \sigma}\cdot\hat{\bm z}_{\text{eff}}$. Note that $\sigma_{{\bm k},\parallel}$ commutes with $H^{(l)}_{0{\bm k}}$, while $\sigma_{{\bm k},k}$ and $\sigma_{{\bm k},z_{\text{eff}}}$ do not, hence we use $\perp$ to refer to vectors in the plane spanned by $\hat{\bm k}$ and $\hat{{\bm z}}_{\text{eff}}$. Hence the kinetic equation for the passive layer Eq.~(\ref{passive}) can be decomposed  as 
\begin{subequations}\label{decomposation}
\begin{equation}\vspace{-0.5cm}
 \frac{\text{d} n^{(\text{p})}_{\bm k}}{\text{d}t}+P_n\hat{J}(f^{(\text{p})}_{\bm k})=\frac{J^{\text{e-e}}_{{\bm k},\text{p},++}+J^{\text{e-e}}_{{\bm k},\text{p},--}}{2}\openone,
\end{equation}
\begin{equation}\label{parallel}\vspace{-0.5cm}
\frac{\text{d}S^{(\text{p})}_{{\bm k},\parallel}}{\text{d}t}+P_{\parallel}\hat{J}(f^{(\text{p})}_{\bm k})=\frac{J^{\text{e-e}}_{{\bm k},\text{p},++}-J^{\text{e-e}}_{{\bm k},\text{p},--}}{2}{\bm \sigma}\cdot\hat{\bm \Omega}_{\bm k},
\end{equation}
\begin{equation}\label{perp}
\ba
&\dps \frac{\text{d}S^{(\text{p})}_{{\bm k},\perp}}{\text{d}t}+\frac{i}{\hbar}[H^{(\text{p})}_{0\bm k},S^{(\text{p})}_{{\bm k},{\perp}}]+P_{\perp}\hat{J}(f^{(\text{p})}_{\bm k})={\bm \sigma}\cdot\hat{\bm k}_{\text{eff}}\\[3ex]
&\dps \times\frac{J^{\text{e-e}}_{{\bm k},\text{p},-+}\!-\!J^{\text{e-e}}_{{\bm k},\text{p},+-}}{2i}\!-\!\frac{J^{,\text{e-e}}_{{\bm k},\text{p},-+}\!+\!J^{\text{e-e}}_{{\bm k},\text{p},+-}}{2}{\bm \sigma}\cdot\hat{\bm z}_{\text{eff}},
\ea
\end{equation}
\end{subequations}
Notation 'e-e' in the formula means interlayer electron-electron scatterings. Solving Eqs.~(\ref{decomposation}) and evaluating the expectation value of the current operator for the passive layer, we find a general form for the longitudinal drag conductivity
\begin{equation}\label{cll}
\sigma^{xx}_\text{D}\!=\!\frac{e^2}{16\pi k_\text{B}T}\sum_{\bm q}\!\int\! \text{d}\omega\frac{|V({\bm q},\omega)|^2\chi_\text{a}({\bm q},
\omega)\chi_\text{p}({\bm q},
\omega)}{\sinh^2 \frac{\beta\hbar\omega}{2}},
\end{equation}
where $\chi_l({\bm q},\omega)$ is the drag susceptibility given in detail in the Supplement. The longitudinal drag conductivity $\sigma^{xx}_D$ was studied in detail in Ref.~\cite{Hong_drag_2015}. In the regime $M_{\text{a,p}} \ll \varepsilon_\text{F}$, which is applicable to all samples studied experimentally, $\sigma^{xx}_D$ does not depend on the magnetization of either layer, and has the same form as in Ref.~\cite{Hong_drag_2015}. 

The anomalous Hall drag current is given by
\begin{equation}\label{jhl-c}
\ba
{\bm j}^{\text{hl}}
&\dps  \!=\!\frac{e^2\pi}{4\hbar k_\text{B}TL^2}\!\int\!\text{d} \omega\frac{|V({\bm q},\omega)|^2}{\sinh^2\frac{\beta\hbar\omega}{2}}\delta[\varepsilon^{(\text{p})}_{k_1,+}\!-\!\varepsilon^{(\text{p})}_{k,+}\!+\!\hbar \omega]\\[3ex]
&\dps \times(f^{(\text{p})}_{0{\bm k},+}\!-\!f^{(\text{p})}_{0{\bm k}_1,+}){\bm E}_{\text{a}}\cdot\chi_\text{a}({\bm q},\omega)\frac{Aa_kb_k}{\Omega_k}(\hat{{\bm \theta}_1}\!-\!\hat{{\bm \theta}}). 
\ea
\end{equation}
Equation~(\ref{jhl-c}) is used to calculate the drag current. It can be written as a function of the Berry curvature of the passive layer, the occupation numbers $(f^{(\text{p})}_{0{\bm k},+}\!-\!f^{(\text{p})}_{0{\bm k}_1,+})$ and an effective driving term due to the electric field and interlayer electron-electron scattering.

Experimentally, for TI films in the large-surface limit, non-topological contributions from the bulk and the side surface are negligible \cite{TME_TI_PRB_Nagaosa}, and we expect the effects described in this work to be observable. Aside from commonly used materials such as Bi$_2$Se$_3$ and Bi$_2$Te$_3$, a small-gap three-dimensional TI has also been identified in strained HgTe. The TI surface states in this material are, however, spatially extended and could be peaked quite far from the wide quantum well edges, reducing the effective 2D layer separation \cite{Tilahun_TIF_QHS_PRL2011}.     

Massive Dirac fermions are also found in graphene with a staggered sublattice potential, and $\text{MoS}_2$ thin films where inversion symmetry is broken. The band Hamiltonian for a single layer is given by $H^{(l)}_{0{\bm k}}=at(\tau k_x\sigma_x+k_y\sigma_y)+\frac{\Delta}{2}\sigma_z$ with $\tau=\pm1$ the valley index, where $a$ is the lattice constant, $t$ is the effective hopping integral, and $\Delta$ is the energy gap. These Dresselhaus-like Hamiltonians can be directly mapped onto the Rashba Hamiltonian considered in this work \cite{DM_Balatsky_2014}. The longitudinal drag current is identical for both TIs and other massive Dirac fermion systems, because the physical mechanism behind the longitudinal drag phenomenon is a result of rectification by the passive layer of the fluctuating electric field generated  by the active layer. However, when inversion symmetry is broken in a 2D hexagonal lattice, a pair of valleys which are time-reversal of each other are distinguishable by their opposite values of magnetic moment and Berry curvature. Therefore, there will be no transverse drag current in graphene or $\text{MoS}_2$ with broken inversion symmetry because the Berry curvatures have opposite signs in the opposite valleys. In a Dirac semimetal, each Dirac point is four-fold degenerate and can be viewed as consisting of two Weyl nodes with opposite chiralities. Consequently, transverse drag currents also vanish in Dirac semimetals. At small magnetizations the longitudinal drag currents in these materials will be independent of the magnetizations of either layer. 
    
The results are also applicable to Rashba 2DEGs, though measuring a strong anomalous Hall effect \cite{Culcer_AHE_PRB03} can be challenging. A sizable fraction of the conductivity quantum is obtained if the two Rashba sub-bands experience a large magnetization splitting and $\varepsilon_\text{F}$ lies in the bottom sub-band, but that is challenging experimentally. When $\varepsilon_\text{F} \gg M$ the effect vanishes altogether. 
    
We have studied Coulomb drag of massive Dirac fermions, finding that the drag due to the topological terms on the active surface vanishes and the only contribution to anomalous Hall drag comes from the anomalous Hall current generated by the longitudinal drag force experienced by the passive layer. In a system in which the active layer is undoped and only a quantized anomalous Hall effect exists in the active layer there is no drag current at all. In a doped system the transverse drag current has a non-monotonic dependence on the magnetization of the passive layer, with a peak at a value of the magnetization that becomes pronounced at lower densities.

The authors thank Di Xiao, Haizhou Lu, A. H. MacDonald, Oleg Sushkov, Tommy Li, and Euyheon Hwang for inspiring discussions.


\begin{thebibliography}{90}
\expandafter\ifx\csname natexlab\endcsname\relax\def\natexlab#1{#1}\fi
\expandafter\ifx\csname bibnamefont\endcsname\relax
  \def\bibnamefont#1{#1}\fi
\expandafter\ifx\csname bibfnamefont\endcsname\relax
  \def\bibfnamefont#1{#1}\fi
\expandafter\ifx\csname citenamefont\endcsname\relax
  \def\citenamefont#1{#1}\fi
\expandafter\ifx\csname url\endcsname\relax
  \def\url#1{\texttt{#1}}\fi
\expandafter\ifx\csname urlprefix\endcsname\relax\def\urlprefix{URL }\fi
\providecommand{\bibinfo}[2]{#2}
\providecommand{\eprint}[2][]{\url{#2}}

\bibitem[{\citenamefont{Kane and Mele}(2005)}]{KaneMele_QSHE_PRL05}
\bibinfo{author}{\bibfnamefont{C.~L.} \bibnamefont{Kane}} \bibnamefont{and}
  \bibinfo{author}{\bibfnamefont{E.~J.} \bibnamefont{Mele}},
  \bibinfo{journal}{Phys.\ Rev.\ Lett.} \textbf{\bibinfo{volume}{95}},
  \bibinfo{pages}{226801} (\bibinfo{year}{2005}).

\bibitem[{\citenamefont{Hasan and Kane}(2010)}]{Hasan_TI_RMP10}
\bibinfo{author}{\bibfnamefont{M.~Z.} \bibnamefont{Hasan}} \bibnamefont{and}
  \bibinfo{author}{\bibfnamefont{C.~L.} \bibnamefont{Kane}},
  \bibinfo{journal}{Rev.\ Mod.\ Phys.} \textbf{\bibinfo{volume}{82}},
  \bibinfo{pages}{3045} (\bibinfo{year}{2010}).

\bibitem[{\citenamefont{Xiao et~al.}(2012)\citenamefont{Xiao, Liu, Feng, Xu,
  and Yao}}]{Dixiao_MoS2_2012}
\bibinfo{author}{\bibfnamefont{D.}~\bibnamefont{Xiao}},
  \bibinfo{author}{\bibfnamefont{G.-B.} \bibnamefont{Liu}},
  \bibinfo{author}{\bibfnamefont{W.}~\bibnamefont{Feng}},
  \bibinfo{author}{\bibfnamefont{X.}~\bibnamefont{Xu}}, \bibnamefont{and}
  \bibinfo{author}{\bibfnamefont{W.}~\bibnamefont{Yao}},
  \bibinfo{journal}{Phys. Rev. Lett.} \textbf{\bibinfo{volume}{108}},
  \bibinfo{pages}{196802} (\bibinfo{year}{2012}).

\bibitem[{\citenamefont{Burkov and Balents}(2011)}]{Burkov_Weyl_TI}
\bibinfo{author}{\bibfnamefont{A.~A.} \bibnamefont{Burkov}} \bibnamefont{and}
  \bibinfo{author}{\bibfnamefont{L.}~\bibnamefont{Balents}},
  \bibinfo{journal}{Phys. Rev. Lett.} \textbf{\bibinfo{volume}{107}},
  \bibinfo{pages}{127205} (\bibinfo{year}{2011}).

\bibitem[{\citenamefont{Edmonds et~al.}(2016)\citenamefont{Edmonds,
  Hellerstedt, O’Donnell, Tadich, and Fuhrer}}]{Fuhrer_Na3Bi_2016}
\bibinfo{author}{\bibfnamefont{M.~T.} \bibnamefont{Edmonds}},
  \bibinfo{author}{\bibfnamefont{J.}~\bibnamefont{Hellerstedt}},
  \bibinfo{author}{\bibfnamefont{K.~M.} \bibnamefont{O’Donnell}},
  \bibinfo{author}{\bibfnamefont{A.}~\bibnamefont{Tadich}}, \bibnamefont{and}
  \bibinfo{author}{\bibfnamefont{M.~S.} \bibnamefont{Fuhrer}},
  \bibinfo{journal}{ACS Applied Materials and Interfaces}
  \textbf{\bibinfo{volume}{8}}, \bibinfo{pages}{16412} (\bibinfo{year}{2016}).

\bibitem[{\citenamefont{Hellerstedt et~al.}(2016)\citenamefont{Hellerstedt,
  Edmonds, Ramakrishnan, Liu, Weber, Tadich, O’Donnell, Adam, and
  Fuhrer}}]{Fuhrer_NL_Dirac_semi}
\bibinfo{author}{\bibfnamefont{J.}~\bibnamefont{Hellerstedt}},
  \bibinfo{author}{\bibfnamefont{M.~T.} \bibnamefont{Edmonds}},
  \bibinfo{author}{\bibfnamefont{N.}~\bibnamefont{Ramakrishnan}},
  \bibinfo{author}{\bibfnamefont{C.}~\bibnamefont{Liu}},
  \bibinfo{author}{\bibfnamefont{B.}~\bibnamefont{Weber}},
  \bibinfo{author}{\bibfnamefont{A.}~\bibnamefont{Tadich}},
  \bibinfo{author}{\bibfnamefont{K.~M.} \bibnamefont{O’Donnell}},
  \bibinfo{author}{\bibfnamefont{S.}~\bibnamefont{Adam}}, \bibnamefont{and}
  \bibinfo{author}{\bibfnamefont{M.~S.} \bibnamefont{Fuhrer}},
  \bibinfo{journal}{Nano Letters} \textbf{\bibinfo{volume}{16}},
  \bibinfo{pages}{3210} (\bibinfo{year}{2016}).

\bibitem[{\citenamefont{Wang et~al.}(2011)\citenamefont{Wang, Zhu, Sun, Li,
  Zhang, Wen, Chen, He, Wang, Ma et~al.}}]{Wang_Bi2Te3_Ctrl_AM11}
\bibinfo{author}{\bibfnamefont{G.}~\bibnamefont{Wang}},
  \bibinfo{author}{\bibfnamefont{X.-G.} \bibnamefont{Zhu}},
  \bibinfo{author}{\bibfnamefont{Y.-Y.} \bibnamefont{Sun}},
  \bibinfo{author}{\bibfnamefont{Y.-Y.} \bibnamefont{Li}},
  \bibinfo{author}{\bibfnamefont{T.}~\bibnamefont{Zhang}},
  \bibinfo{author}{\bibfnamefont{J.}~\bibnamefont{Wen}},
  \bibinfo{author}{\bibfnamefont{X.}~\bibnamefont{Chen}},
  \bibinfo{author}{\bibfnamefont{K.}~\bibnamefont{He}},
  \bibinfo{author}{\bibfnamefont{L.-L.} \bibnamefont{Wang}},
  \bibinfo{author}{\bibfnamefont{X.-C.} \bibnamefont{Ma}},
  \bibnamefont{et~al.}, \bibinfo{journal}{Advanced Materials}
  \textbf{\bibinfo{volume}{23}}, \bibinfo{pages}{2929} (\bibinfo{year}{2011}).

\bibitem[{\citenamefont{Sac{\'e}p{\'e}
  et~al.}(2011)\citenamefont{Sac{\'e}p{\'e}, Oostinga, Li, Ubaldini, Couto,
  Giannini, and Morpurgo}}]{Benjamin_nature_2011}
\bibinfo{author}{\bibfnamefont{B.}~\bibnamefont{Sac{\'e}p{\'e}}},
  \bibinfo{author}{\bibfnamefont{J.~B.} \bibnamefont{Oostinga}},
  \bibinfo{author}{\bibfnamefont{J.}~\bibnamefont{Li}},
  \bibinfo{author}{\bibfnamefont{A.}~\bibnamefont{Ubaldini}},
  \bibinfo{author}{\bibfnamefont{N.~J.~G.} \bibnamefont{Couto}},
  \bibinfo{author}{\bibfnamefont{E.}~\bibnamefont{Giannini}}, \bibnamefont{and}
  \bibinfo{author}{\bibfnamefont{A.~F.} \bibnamefont{Morpurgo}},
  \bibinfo{journal}{Nat. Commun.} \textbf{\bibinfo{volume}{2}},
  \bibinfo{pages}{575} (\bibinfo{year}{2011}).

\bibitem[{\citenamefont{He et~al.}(2011)\citenamefont{He, Wang, Zhang, Sou,
  Wong, Wang, Lu, Shen, and Zhang}}]{He_Bi2Te3_Film_WAL_ImpEff_PRL11}
\bibinfo{author}{\bibfnamefont{H.-T.} \bibnamefont{He}},
  \bibinfo{author}{\bibfnamefont{G.}~\bibnamefont{Wang}},
  \bibinfo{author}{\bibfnamefont{T.}~\bibnamefont{Zhang}},
  \bibinfo{author}{\bibfnamefont{I.-K.} \bibnamefont{Sou}},
  \bibinfo{author}{\bibfnamefont{G.~K.~L.} \bibnamefont{Wong}},
  \bibinfo{author}{\bibfnamefont{J.-N.} \bibnamefont{Wang}},
  \bibinfo{author}{\bibfnamefont{H.-Z.} \bibnamefont{Lu}},
  \bibinfo{author}{\bibfnamefont{S.-Q.} \bibnamefont{Shen}}, \bibnamefont{and}
  \bibinfo{author}{\bibfnamefont{F.-C.} \bibnamefont{Zhang}},
  \bibinfo{journal}{Phys. Rev. Lett.} \textbf{\bibinfo{volume}{106}},
  \bibinfo{pages}{166805} (\bibinfo{year}{2011}).

\bibitem[{\citenamefont{Ren et~al.}(2012)\citenamefont{Ren, Taskin, Sasaki,
  Segawa, and Ando}}]{Ren-PRB-2012}
\bibinfo{author}{\bibfnamefont{Z.}~\bibnamefont{Ren}},
  \bibinfo{author}{\bibfnamefont{A.~A.} \bibnamefont{Taskin}},
  \bibinfo{author}{\bibfnamefont{S.}~\bibnamefont{Sasaki}},
  \bibinfo{author}{\bibfnamefont{K.}~\bibnamefont{Segawa}}, \bibnamefont{and}
  \bibinfo{author}{\bibfnamefont{Y.}~\bibnamefont{Ando}},
  \bibinfo{journal}{Phys. Rev. B} \textbf{\bibinfo{volume}{85}},
  \bibinfo{pages}{155301} (\bibinfo{year}{2012}).

\bibitem[{\citenamefont{Kim et~al.}(2012)\citenamefont{Kim, Cho, Butch, Syers,
  Kirshenbaum, Adam, Paglione, and Fuhrer}}]{Kim_sur_2012}
\bibinfo{author}{\bibfnamefont{D.}~\bibnamefont{Kim}},
  \bibinfo{author}{\bibfnamefont{S.}~\bibnamefont{Cho}},
  \bibinfo{author}{\bibfnamefont{N.~P.} \bibnamefont{Butch}},
  \bibinfo{author}{\bibfnamefont{P.}~\bibnamefont{Syers}},
  \bibinfo{author}{\bibfnamefont{K.}~\bibnamefont{Kirshenbaum}},
  \bibinfo{author}{\bibfnamefont{S.}~\bibnamefont{Adam}},
  \bibinfo{author}{\bibfnamefont{J.}~\bibnamefont{Paglione}}, \bibnamefont{and}
  \bibinfo{author}{\bibfnamefont{M.~S.} \bibnamefont{Fuhrer}},
  \bibinfo{journal}{Nat. Phys.} \textbf{\bibinfo{volume}{8}},
  \bibinfo{pages}{459} (\bibinfo{year}{2012}).

\bibitem[{\citenamefont{Zhang et~al.}(2013)\citenamefont{Zhang, Chang, Tang,
  Zhang, Feng, Li, Wang, Chen, Liu, Duan
  et~al.}}]{Jinsong-Zhang-Yayu-Wang-2013-science}
\bibinfo{author}{\bibfnamefont{J.}~\bibnamefont{Zhang}},
  \bibinfo{author}{\bibfnamefont{C.-Z.} \bibnamefont{Chang}},
  \bibinfo{author}{\bibfnamefont{P.}~\bibnamefont{Tang}},
  \bibinfo{author}{\bibfnamefont{Z.}~\bibnamefont{Zhang}},
  \bibinfo{author}{\bibfnamefont{X.}~\bibnamefont{Feng}},
  \bibinfo{author}{\bibfnamefont{K.}~\bibnamefont{Li}},
  \bibinfo{author}{\bibfnamefont{L.-l.} \bibnamefont{Wang}},
  \bibinfo{author}{\bibfnamefont{X.}~\bibnamefont{Chen}},
  \bibinfo{author}{\bibfnamefont{C.}~\bibnamefont{Liu}},
  \bibinfo{author}{\bibfnamefont{W.}~\bibnamefont{Duan}}, \bibnamefont{et~al.},
  \bibinfo{journal}{Science} \textbf{\bibinfo{volume}{339}},
  \bibinfo{pages}{15821586} (\bibinfo{year}{2013}).

\bibitem[{\citenamefont{Barreto et~al.}(2014)\citenamefont{Barreto, Khnemund,
  Edler, Tegenkamp, Mi, Bremholm, Iversen, Frydendahl, Bianchi, and
  Hofmann}}]{Lucas_nl_tran_2014}
\bibinfo{author}{\bibfnamefont{L.}~\bibnamefont{Barreto}},
  \bibinfo{author}{\bibfnamefont{L.}~\bibnamefont{Khnemund}},
  \bibinfo{author}{\bibfnamefont{F.}~\bibnamefont{Edler}},
  \bibinfo{author}{\bibfnamefont{C.}~\bibnamefont{Tegenkamp}},
  \bibinfo{author}{\bibfnamefont{J.}~\bibnamefont{Mi}},
  \bibinfo{author}{\bibfnamefont{M.}~\bibnamefont{Bremholm}},
  \bibinfo{author}{\bibfnamefont{B.~B.} \bibnamefont{Iversen}},
  \bibinfo{author}{\bibfnamefont{C.}~\bibnamefont{Frydendahl}},
  \bibinfo{author}{\bibfnamefont{M.}~\bibnamefont{Bianchi}}, \bibnamefont{and}
  \bibinfo{author}{\bibfnamefont{P.}~\bibnamefont{Hofmann}},
  \bibinfo{journal}{Nano Letters} \textbf{\bibinfo{volume}{14}},
  \bibinfo{pages}{3755} (\bibinfo{year}{2014}).

\bibitem[{\citenamefont{Tang et~al.}(2014)\citenamefont{Tang, Chang, Kou,
  Murata, Choi, Lang, Fan, Jiang, Montazeri, Jiang
  et~al.}}]{Jianshi_nl_spin_polarized_2014}
\bibinfo{author}{\bibfnamefont{J.}~\bibnamefont{Tang}},
  \bibinfo{author}{\bibfnamefont{L.-T.} \bibnamefont{Chang}},
  \bibinfo{author}{\bibfnamefont{X.}~\bibnamefont{Kou}},
  \bibinfo{author}{\bibfnamefont{K.}~\bibnamefont{Murata}},
  \bibinfo{author}{\bibfnamefont{E.~S.} \bibnamefont{Choi}},
  \bibinfo{author}{\bibfnamefont{M.}~\bibnamefont{Lang}},
  \bibinfo{author}{\bibfnamefont{Y.}~\bibnamefont{Fan}},
  \bibinfo{author}{\bibfnamefont{Y.}~\bibnamefont{Jiang}},
  \bibinfo{author}{\bibfnamefont{M.}~\bibnamefont{Montazeri}},
  \bibinfo{author}{\bibfnamefont{W.}~\bibnamefont{Jiang}},
  \bibnamefont{et~al.}, \bibinfo{journal}{Nano Letters}
  \textbf{\bibinfo{volume}{14}}, \bibinfo{pages}{5423} (\bibinfo{year}{2014}).

\bibitem[{\citenamefont{Kozlov et~al.}(2014)\citenamefont{Kozlov, Kvon,
  Olshanetsky, Mikhailov, Dvoretsky, and Weiss}}]{Kozlov_prl_2014}
\bibinfo{author}{\bibfnamefont{D.~A.} \bibnamefont{Kozlov}},
  \bibinfo{author}{\bibfnamefont{Z.~D.} \bibnamefont{Kvon}},
  \bibinfo{author}{\bibfnamefont{E.~B.} \bibnamefont{Olshanetsky}},
  \bibinfo{author}{\bibfnamefont{N.~N.} \bibnamefont{Mikhailov}},
  \bibinfo{author}{\bibfnamefont{S.~A.} \bibnamefont{Dvoretsky}},
  \bibnamefont{and} \bibinfo{author}{\bibfnamefont{D.}~\bibnamefont{Weiss}},
  \bibinfo{journal}{Phys. Rev. Lett.} \textbf{\bibinfo{volume}{112}},
  \bibinfo{pages}{196801} (\bibinfo{year}{2014}).

\bibitem[{\citenamefont{Kim et~al.}(2013)\citenamefont{Kim, Syers, Butch,
  Paglione, and Fuhrer}}]{Fuhrer_nature_2013}
\bibinfo{author}{\bibfnamefont{D.}~\bibnamefont{Kim}},
  \bibinfo{author}{\bibfnamefont{P.}~\bibnamefont{Syers}},
  \bibinfo{author}{\bibfnamefont{N.~P.} \bibnamefont{Butch}},
  \bibinfo{author}{\bibfnamefont{J.}~\bibnamefont{Paglione}}, \bibnamefont{and}
  \bibinfo{author}{\bibfnamefont{M.~S.} \bibnamefont{Fuhrer}},
  \bibinfo{journal}{Nat. Commun.} \textbf{\bibinfo{volume}{4}}
  (\bibinfo{year}{2013}).

\bibitem[{\citenamefont{Cacho et~al.}(2015{\natexlab{a}})\citenamefont{Cacho,
  Crepaldi, Battiato, Braun, Cilento, Zacchigna, Richter, Heckmann, Springate,
  Liu et~al.}}]{Cacho_prl_2015}
\bibinfo{author}{\bibfnamefont{C.}~\bibnamefont{Cacho}},
  \bibinfo{author}{\bibfnamefont{A.}~\bibnamefont{Crepaldi}},
  \bibinfo{author}{\bibfnamefont{M.}~\bibnamefont{Battiato}},
  \bibinfo{author}{\bibfnamefont{J.}~\bibnamefont{Braun}},
  \bibinfo{author}{\bibfnamefont{F.}~\bibnamefont{Cilento}},
  \bibinfo{author}{\bibfnamefont{M.}~\bibnamefont{Zacchigna}},
  \bibinfo{author}{\bibfnamefont{M.}~\bibnamefont{Richter}},
  \bibinfo{author}{\bibfnamefont{O.}~\bibnamefont{Heckmann}},
  \bibinfo{author}{\bibfnamefont{E.}~\bibnamefont{Springate}},
  \bibinfo{author}{\bibfnamefont{Y.}~\bibnamefont{Liu}}, \bibnamefont{et~al.},
  \bibinfo{journal}{Phys. Rev. Lett.} \textbf{\bibinfo{volume}{114}},
  \bibinfo{pages}{097401} (\bibinfo{year}{2015}{\natexlab{a}}).

\bibitem[{\citenamefont{Zhang et~al.}(2015)\citenamefont{Zhang, Feng, Xu, Guo,
  Zhang, Ou, Feng, Li, Zhang, Wang et~al.}}]{ZhangJinsong_prb_2015}
\bibinfo{author}{\bibfnamefont{J.}~\bibnamefont{Zhang}},
  \bibinfo{author}{\bibfnamefont{X.}~\bibnamefont{Feng}},
  \bibinfo{author}{\bibfnamefont{Y.}~\bibnamefont{Xu}},
  \bibinfo{author}{\bibfnamefont{M.}~\bibnamefont{Guo}},
  \bibinfo{author}{\bibfnamefont{Z.}~\bibnamefont{Zhang}},
  \bibinfo{author}{\bibfnamefont{Y.}~\bibnamefont{Ou}},
  \bibinfo{author}{\bibfnamefont{Y.}~\bibnamefont{Feng}},
  \bibinfo{author}{\bibfnamefont{K.}~\bibnamefont{Li}},
  \bibinfo{author}{\bibfnamefont{H.}~\bibnamefont{Zhang}},
  \bibinfo{author}{\bibfnamefont{L.}~\bibnamefont{Wang}}, \bibnamefont{et~al.},
  \bibinfo{journal}{Phys. Rev. B} \textbf{\bibinfo{volume}{91}},
  \bibinfo{pages}{075431} (\bibinfo{year}{2015}).

\bibitem[{\citenamefont{Hellerstedt et~al.}(2014)\citenamefont{Hellerstedt,
  Edmonds, Chen, Cullen, Zheng, and Fuhrer}}]{Hellerstedt_APL_2014}
\bibinfo{author}{\bibfnamefont{J.}~\bibnamefont{Hellerstedt}},
  \bibinfo{author}{\bibfnamefont{M.~T.} \bibnamefont{Edmonds}},
  \bibinfo{author}{\bibfnamefont{J.~H.} \bibnamefont{Chen}},
  \bibinfo{author}{\bibfnamefont{W.~G.} \bibnamefont{Cullen}},
  \bibinfo{author}{\bibfnamefont{C.~X.} \bibnamefont{Zheng}}, \bibnamefont{and}
  \bibinfo{author}{\bibfnamefont{M.~S.} \bibnamefont{Fuhrer}},
  \bibinfo{journal}{Applied Physics Letters} \textbf{\bibinfo{volume}{105}},
  \bibinfo{pages}{173506} (\bibinfo{year}{2014}).

\bibitem[{\citenamefont{Kastl et~al.}(2015)\citenamefont{Kastl, Karnetzky,
  Karl, and Holleitner}}]{Kastl_surface_tran_nature_2015}
\bibinfo{author}{\bibfnamefont{C.}~\bibnamefont{Kastl}},
  \bibinfo{author}{\bibfnamefont{C.}~\bibnamefont{Karnetzky}},
  \bibinfo{author}{\bibfnamefont{H.}~\bibnamefont{Karl}}, \bibnamefont{and}
  \bibinfo{author}{\bibfnamefont{A.~W.} \bibnamefont{Holleitner}},
  \bibinfo{journal}{Nat. Commun.} \textbf{\bibinfo{volume}{6}}
  (\bibinfo{year}{2015}).

\bibitem[{\citenamefont{Fan et~al.}(2016)\citenamefont{Fan, Kou, Upadhyaya,
  Shao, Pan, Lang, Che, Tang, Montazeri, Murata et~al.}}]{E_control_SOT_TI}
\bibinfo{author}{\bibfnamefont{Y.}~\bibnamefont{Fan}},
  \bibinfo{author}{\bibfnamefont{X.}~\bibnamefont{Kou}},
  \bibinfo{author}{\bibfnamefont{P.}~\bibnamefont{Upadhyaya}},
  \bibinfo{author}{\bibfnamefont{Q.}~\bibnamefont{Shao}},
  \bibinfo{author}{\bibfnamefont{L.}~\bibnamefont{Pan}},
  \bibinfo{author}{\bibfnamefont{M.}~\bibnamefont{Lang}},
  \bibinfo{author}{\bibfnamefont{X.}~\bibnamefont{Che}},
  \bibinfo{author}{\bibfnamefont{J.}~\bibnamefont{Tang}},
  \bibinfo{author}{\bibfnamefont{M.}~\bibnamefont{Montazeri}},
  \bibinfo{author}{\bibfnamefont{K.}~\bibnamefont{Murata}},
  \bibnamefont{et~al.}, \bibinfo{journal}{Nat Nano}
  \textbf{\bibinfo{volume}{11}}, \bibinfo{pages}{352359}
  (\bibinfo{year}{2016}).

\bibitem[{\citenamefont{Hwang and {Das
  Sarma}}(2007)}]{Hwang_Gfn_Screening_PRB07}
\bibinfo{author}{\bibfnamefont{E.~H.} \bibnamefont{Hwang}} \bibnamefont{and}
  \bibinfo{author}{\bibfnamefont{S.}~\bibnamefont{{Das Sarma}}},
  \bibinfo{journal}{Phys.\ Rev.\ B} \textbf{\bibinfo{volume}{75}},
  \bibinfo{pages}{205418} (\bibinfo{year}{2007}).

\bibitem[{\citenamefont{Jung and
  MacDonald}(2011)}]{JungMacDonald_graphene_PRB2011}
\bibinfo{author}{\bibfnamefont{J.}~\bibnamefont{Jung}} \bibnamefont{and}
  \bibinfo{author}{\bibfnamefont{A.~H.} \bibnamefont{MacDonald}},
  \bibinfo{journal}{Phys. Rev. B} \textbf{\bibinfo{volume}{84}},
  \bibinfo{pages}{085446} (\bibinfo{year}{2011}).

\bibitem[{\citenamefont{Durst}(2015)}]{Durst_2015}
\bibinfo{author}{\bibfnamefont{A.~C.} \bibnamefont{Durst}},
  \bibinfo{journal}{Phys. Rev. B} \textbf{\bibinfo{volume}{91}},
  \bibinfo{pages}{094519} (\bibinfo{year}{2015}).

\bibitem[{\citenamefont{Lu and Shen}(2014{\natexlab{a}})}]{Haizhou_WL_2014}
\bibinfo{author}{\bibfnamefont{H.-Z.} \bibnamefont{Lu}} \bibnamefont{and}
  \bibinfo{author}{\bibfnamefont{S.-Q.} \bibnamefont{Shen}},
  \bibinfo{journal}{Proc. SPIE} \textbf{\bibinfo{volume}{9167}},
  \bibinfo{pages}{91672E} (\bibinfo{year}{2014}{\natexlab{a}}).

\bibitem[{\citenamefont{Zhang et~al.}(2010)\citenamefont{Zhang, Yu, Zhang, Dai,
  and Fang}}]{FP_Bi2X_3_NewJ}
\bibinfo{author}{\bibfnamefont{W.}~\bibnamefont{Zhang}},
  \bibinfo{author}{\bibfnamefont{R.}~\bibnamefont{Yu}},
  \bibinfo{author}{\bibfnamefont{H.-J.} \bibnamefont{Zhang}},
  \bibinfo{author}{\bibfnamefont{X.}~\bibnamefont{Dai}}, \bibnamefont{and}
  \bibinfo{author}{\bibfnamefont{Z.}~\bibnamefont{Fang}}, \bibinfo{journal}{New
  Journal of Physics} \textbf{\bibinfo{volume}{12}}, \bibinfo{pages}{065013}
  (\bibinfo{year}{2010}).

\bibitem[{\citenamefont{Liu et~al.}(2016{\natexlab{a}})\citenamefont{Liu, Kim,
  Tan, and Rappe}}]{Shi_Rappe_nl_2016}
\bibinfo{author}{\bibfnamefont{S.}~\bibnamefont{Liu}},
  \bibinfo{author}{\bibfnamefont{Y.}~\bibnamefont{Kim}},
  \bibinfo{author}{\bibfnamefont{L.~Z.} \bibnamefont{Tan}}, \bibnamefont{and}
  \bibinfo{author}{\bibfnamefont{A.~M.} \bibnamefont{Rappe}},
  \bibinfo{journal}{Nano Letters} \textbf{\bibinfo{volume}{16}},
  \bibinfo{pages}{16631668} (\bibinfo{year}{2016}{\natexlab{a}}).

\bibitem[{\citenamefont{Culcer and Das~Sarma}(2011)}]{Culcer_TI_AHE_PRB11}
\bibinfo{author}{\bibfnamefont{D.}~\bibnamefont{Culcer}} \bibnamefont{and}
  \bibinfo{author}{\bibfnamefont{S.}~\bibnamefont{Das~Sarma}},
  \bibinfo{journal}{Phys. Rev. B} \textbf{\bibinfo{volume}{83}},
  \bibinfo{pages}{245441} (\bibinfo{year}{2011}).

\bibitem[{\citenamefont{Adam et~al.}(2012)\citenamefont{Adam, Hwang, and
  Das~Sarma}}]{Adam_2D_Tran_prb_2012}
\bibinfo{author}{\bibfnamefont{S.}~\bibnamefont{Adam}},
  \bibinfo{author}{\bibfnamefont{E.~H.} \bibnamefont{Hwang}}, \bibnamefont{and}
  \bibinfo{author}{\bibfnamefont{S.}~\bibnamefont{Das~Sarma}},
  \bibinfo{journal}{Phys. Rev. B} \textbf{\bibinfo{volume}{85}},
  \bibinfo{pages}{235413} (\bibinfo{year}{2012}).

\bibitem[{\citenamefont{Yoshida et~al.}(2012)\citenamefont{Yoshida, Fujimoto,
  and Kawakami}}]{Yoshida-PRB-2012}
\bibinfo{author}{\bibfnamefont{T.}~\bibnamefont{Yoshida}},
  \bibinfo{author}{\bibfnamefont{S.}~\bibnamefont{Fujimoto}}, \bibnamefont{and}
  \bibinfo{author}{\bibfnamefont{N.}~\bibnamefont{Kawakami}},
  \bibinfo{journal}{Phys. Rev. B} \textbf{\bibinfo{volume}{85}},
  \bibinfo{pages}{125113} (\bibinfo{year}{2012}).

\bibitem[{\citenamefont{Das~Sarma and Li}(2013)}]{LiQiuzi_2013_prb}
\bibinfo{author}{\bibfnamefont{S.}~\bibnamefont{Das~Sarma}} \bibnamefont{and}
  \bibinfo{author}{\bibfnamefont{Q.}~\bibnamefont{Li}}, \bibinfo{journal}{Phys.
  Rev. B} \textbf{\bibinfo{volume}{88}}, \bibinfo{pages}{081404}
  (\bibinfo{year}{2013}).

\bibitem[{\citenamefont{Liu et~al.}(2014)\citenamefont{Liu, Liu, and
  Culcer}}]{Weizhe_TITF_2014_prb}
\bibinfo{author}{\bibfnamefont{W.~E.} \bibnamefont{Liu}},
  \bibinfo{author}{\bibfnamefont{H.}~\bibnamefont{Liu}}, \bibnamefont{and}
  \bibinfo{author}{\bibfnamefont{D.}~\bibnamefont{Culcer}},
  \bibinfo{journal}{Phys. Rev. B} \textbf{\bibinfo{volume}{89}},
  \bibinfo{pages}{195417} (\bibinfo{year}{2014}).

\bibitem[{\citenamefont{Lu and
  Shen}(2014{\natexlab{b}})}]{Hai-Zhou_Conductivity_2014_prl}
\bibinfo{author}{\bibfnamefont{H.-Z.} \bibnamefont{Lu}} \bibnamefont{and}
  \bibinfo{author}{\bibfnamefont{S.-Q.} \bibnamefont{Shen}},
  \bibinfo{journal}{Phys. Rev. Lett.} \textbf{\bibinfo{volume}{112}},
  \bibinfo{pages}{146601} (\bibinfo{year}{2014}{\natexlab{b}}).

\bibitem[{\citenamefont{Das~Sarma et~al.}(2015)\citenamefont{Das~Sarma, Hwang,
  and Min}}]{Das_prb_2015}
\bibinfo{author}{\bibfnamefont{S.}~\bibnamefont{Das~Sarma}},
  \bibinfo{author}{\bibfnamefont{E.~H.} \bibnamefont{Hwang}}, \bibnamefont{and}
  \bibinfo{author}{\bibfnamefont{H.}~\bibnamefont{Min}},
  \bibinfo{journal}{Phys. Rev. B} \textbf{\bibinfo{volume}{91}},
  \bibinfo{pages}{035201} (\bibinfo{year}{2015}).

\bibitem[{\citenamefont{Nayak et~al.}(2008)\citenamefont{Nayak, Simon, Stern,
  Freedman, and Das~Sarma}}]{Das_Sarma_RMP_TQC}
\bibinfo{author}{\bibfnamefont{C.}~\bibnamefont{Nayak}},
  \bibinfo{author}{\bibfnamefont{S.~H.} \bibnamefont{Simon}},
  \bibinfo{author}{\bibfnamefont{A.}~\bibnamefont{Stern}},
  \bibinfo{author}{\bibfnamefont{M.}~\bibnamefont{Freedman}}, \bibnamefont{and}
  \bibinfo{author}{\bibfnamefont{S.}~\bibnamefont{Das~Sarma}},
  \bibinfo{journal}{Rev. Mod. Phys.} \textbf{\bibinfo{volume}{80}},
  \bibinfo{pages}{10831159} (\bibinfo{year}{2008}).

\bibitem[{\citenamefont{Pograbinskii}(1977)}]{1977_drag}
\bibinfo{author}{\bibfnamefont{M.}~\bibnamefont{Pograbinskii}},
  \bibinfo{journal}{Sov.Phys.Semicond.} \textbf{\bibinfo{volume}{11}},
  \bibinfo{pages}{372} (\bibinfo{year}{1977}).

\bibitem[{\citenamefont{Zheng and MacDonald}(1993)}]{Zheng-2DES-1993}
\bibinfo{author}{\bibfnamefont{L.}~\bibnamefont{Zheng}} \bibnamefont{and}
  \bibinfo{author}{\bibfnamefont{A.~H.} \bibnamefont{MacDonald}},
  \bibinfo{journal}{Phys. Rev. B} \textbf{\bibinfo{volume}{48}},
  \bibinfo{pages}{8203} (\bibinfo{year}{1993}).

\bibitem[{\citenamefont{Narozhny and Levchenko}(to be published)}]{Drag_riview}
\bibinfo{author}{\bibfnamefont{B.}~\bibnamefont{Narozhny}} \bibnamefont{and}
  \bibinfo{author}{\bibfnamefont{A.}~\bibnamefont{Levchenko}},
  \bibinfo{journal}{arXiv:1505.07468}  (\bibinfo{year}{to be published}).

\bibitem[{\citenamefont{Kaasbjerg and Jauho}(to be published)}]{Drag-dot}
\bibinfo{author}{\bibfnamefont{K.}~\bibnamefont{Kaasbjerg}} \bibnamefont{and}
  \bibinfo{author}{\bibfnamefont{A.-P.} \bibnamefont{Jauho}},
  \bibinfo{journal}{arXiv:1601.00673}  (\bibinfo{year}{to be published}).

\bibitem[{\citenamefont{D et~al.}(2014)\citenamefont{D, G, Lilly, and
  Reno}}]{1D-1D_science_Luttinger}
\bibinfo{author}{\bibfnamefont{L.}~\bibnamefont{D}},
  \bibinfo{author}{\bibfnamefont{G.}~\bibnamefont{G}},
  \bibinfo{author}{\bibfnamefont{M.~P.} \bibnamefont{Lilly}}, \bibnamefont{and}
  \bibinfo{author}{\bibfnamefont{J.~L.} \bibnamefont{Reno}},
  \bibinfo{journal}{Science} \textbf{\bibinfo{volume}{343}},
  \bibinfo{pages}{631634} (\bibinfo{year}{2014}).

\bibitem[{\citenamefont{Dmitriev et~al.}(2012)\citenamefont{Dmitriev, Gornyi,
  and Polyakov}}]{1D-1D_prb_Dmitriev}
\bibinfo{author}{\bibfnamefont{A.~P.} \bibnamefont{Dmitriev}},
  \bibinfo{author}{\bibfnamefont{I.~V.} \bibnamefont{Gornyi}},
  \bibnamefont{and} \bibinfo{author}{\bibfnamefont{D.~G.}
  \bibnamefont{Polyakov}}, \bibinfo{journal}{Phys. Rev. B}
  \textbf{\bibinfo{volume}{86}}, \bibinfo{pages}{245402}
  (\bibinfo{year}{2012}).

\bibitem[{\citenamefont{Zyuzin and Fiete}(2010)}]{edge_drag_prb}
\bibinfo{author}{\bibfnamefont{V.~A.} \bibnamefont{Zyuzin}} \bibnamefont{and}
  \bibinfo{author}{\bibfnamefont{G.~A.} \bibnamefont{Fiete}},
  \bibinfo{journal}{Phys. Rev. B} \textbf{\bibinfo{volume}{82}},
  \bibinfo{pages}{113305} (\bibinfo{year}{2010}).

\bibitem[{\citenamefont{Jauho and Smith}(1993)}]{ee_drag_93_prb}
\bibinfo{author}{\bibfnamefont{A.-P.} \bibnamefont{Jauho}} \bibnamefont{and}
  \bibinfo{author}{\bibfnamefont{H.}~\bibnamefont{Smith}},
  \bibinfo{journal}{Phys. Rev. B} \textbf{\bibinfo{volume}{47}},
  \bibinfo{pages}{4420} (\bibinfo{year}{1993}).

\bibitem[{\citenamefont{Nandi et~al.}(2012)\citenamefont{Nandi, Finck,
  Eisenstein, Pfeiffer, and West}}]{Nandi_eh_nat}
\bibinfo{author}{\bibfnamefont{D.}~\bibnamefont{Nandi}},
  \bibinfo{author}{\bibfnamefont{A.~D.~K.} \bibnamefont{Finck}},
  \bibinfo{author}{\bibfnamefont{J.~P.} \bibnamefont{Eisenstein}},
  \bibinfo{author}{\bibfnamefont{L.~N.} \bibnamefont{Pfeiffer}},
  \bibnamefont{and} \bibinfo{author}{\bibfnamefont{K.~W.} \bibnamefont{West}},
  \bibinfo{journal}{Nature} \textbf{\bibinfo{volume}{488}},
  \bibinfo{pages}{481484} (\bibinfo{year}{2012}).

\bibitem[{\citenamefont{Efimkin and Galitski}(2016)}]{E-H_exciton_2016}
\bibinfo{author}{\bibfnamefont{D.~K.} \bibnamefont{Efimkin}} \bibnamefont{and}
  \bibinfo{author}{\bibfnamefont{V.}~\bibnamefont{Galitski}},
  \bibinfo{journal}{Phys. Rev. Lett.} \textbf{\bibinfo{volume}{116}},
  \bibinfo{pages}{046801} (\bibinfo{year}{2016}).

\bibitem[{\citenamefont{Gornyi et~al.}(1999)\citenamefont{Gornyi, Yashenkin,
  and Khveshchenko}}]{Disorder_drag}
\bibinfo{author}{\bibfnamefont{I.~V.} \bibnamefont{Gornyi}},
  \bibinfo{author}{\bibfnamefont{A.~G.} \bibnamefont{Yashenkin}},
  \bibnamefont{and} \bibinfo{author}{\bibfnamefont{D.~V.}
  \bibnamefont{Khveshchenko}}, \bibinfo{journal}{Phys. Rev. Lett.}
  \textbf{\bibinfo{volume}{83}}, \bibinfo{pages}{152} (\bibinfo{year}{1999}).

\bibitem[{\citenamefont{Jauho}(1996)}]{Drag_semi_1996}
\bibinfo{author}{\bibfnamefont{A.~P.} \bibnamefont{Jauho}},
  \bibinfo{journal}{IEEE} \textbf{\bibinfo{volume}{1}}, \bibinfo{pages}{2130}
  (\bibinfo{year}{1996}).

\bibitem[{\citenamefont{Pillarisetty et~al.}(2002)\citenamefont{Pillarisetty,
  Noh, Tsui, De~Poortere, Tutuc, and Shayegan}}]{Semi_drag_prl}
\bibinfo{author}{\bibfnamefont{R.}~\bibnamefont{Pillarisetty}},
  \bibinfo{author}{\bibfnamefont{H.}~\bibnamefont{Noh}},
  \bibinfo{author}{\bibfnamefont{D.~C.} \bibnamefont{Tsui}},
  \bibinfo{author}{\bibfnamefont{E.~P.} \bibnamefont{De~Poortere}},
  \bibinfo{author}{\bibfnamefont{E.}~\bibnamefont{Tutuc}}, \bibnamefont{and}
  \bibinfo{author}{\bibfnamefont{M.}~\bibnamefont{Shayegan}},
  \bibinfo{journal}{Phys. Rev. Lett.} \textbf{\bibinfo{volume}{89}},
  \bibinfo{pages}{016805} (\bibinfo{year}{2002}).

\bibitem[{\citenamefont{Kamenev and Oreg}(1995)}]{1995_Hall_drag}
\bibinfo{author}{\bibfnamefont{A.}~\bibnamefont{Kamenev}} \bibnamefont{and}
  \bibinfo{author}{\bibfnamefont{Y.}~\bibnamefont{Oreg}},
  \bibinfo{journal}{Phys. Rev. B} \textbf{\bibinfo{volume}{52}},
  \bibinfo{pages}{7516} (\bibinfo{year}{1995}).

\bibitem[{\citenamefont{Braude and Stern}(2001)}]{drag_metal_prb}
\bibinfo{author}{\bibfnamefont{V.}~\bibnamefont{Braude}} \bibnamefont{and}
  \bibinfo{author}{\bibfnamefont{A.}~\bibnamefont{Stern}},
  \bibinfo{journal}{Phys. Rev. B} \textbf{\bibinfo{volume}{64}},
  \bibinfo{pages}{115431} (\bibinfo{year}{2001}).

\bibitem[{\citenamefont{Tse et~al.}(2007)\citenamefont{Tse, Hu, and
  Das~Sarma}}]{Tse2007}
\bibinfo{author}{\bibfnamefont{W.-K.} \bibnamefont{Tse}},
  \bibinfo{author}{\bibfnamefont{B.~Y.-K.} \bibnamefont{Hu}}, \bibnamefont{and}
  \bibinfo{author}{\bibfnamefont{S.}~\bibnamefont{Das~Sarma}},
  \bibinfo{journal}{Phys. Rev. B} \textbf{\bibinfo{volume}{76}},
  \bibinfo{pages}{081401} (\bibinfo{year}{2007}).

\bibitem[{\citenamefont{Amorim and Peres}(2012)}]{Amorim_drag}
\bibinfo{author}{\bibfnamefont{B.}~\bibnamefont{Amorim}} \bibnamefont{and}
  \bibinfo{author}{\bibfnamefont{N.~M.~R.} \bibnamefont{Peres}},
  \bibinfo{journal}{Journal of Physics: Condensed Matter}
  \textbf{\bibinfo{volume}{24}}, \bibinfo{pages}{335602}
  (\bibinfo{year}{2012}).

\bibitem[{\citenamefont{Kim et~al.}(2011)\citenamefont{Kim, Jo, Nah, Yao,
  Banerjee, and Tutuc}}]{Kim2011}
\bibinfo{author}{\bibfnamefont{S.}~\bibnamefont{Kim}},
  \bibinfo{author}{\bibfnamefont{I.}~\bibnamefont{Jo}},
  \bibinfo{author}{\bibfnamefont{J.}~\bibnamefont{Nah}},
  \bibinfo{author}{\bibfnamefont{Z.}~\bibnamefont{Yao}},
  \bibinfo{author}{\bibfnamefont{S.~K.} \bibnamefont{Banerjee}},
  \bibnamefont{and} \bibinfo{author}{\bibfnamefont{E.}~\bibnamefont{Tutuc}},
  \bibinfo{journal}{Phys. Rev. B} \textbf{\bibinfo{volume}{83}},
  \bibinfo{pages}{161401} (\bibinfo{year}{2011}).

\bibitem[{\citenamefont{Tse and Das~Sarma}(2007)}]{Tse_SO_Drag_PRB07}
\bibinfo{author}{\bibfnamefont{W.-K.} \bibnamefont{Tse}} \bibnamefont{and}
  \bibinfo{author}{\bibfnamefont{S.}~\bibnamefont{Das~Sarma}},
  \bibinfo{journal}{Phys. Rev. B} \textbf{\bibinfo{volume}{75}},
  \bibinfo{pages}{045333} (\bibinfo{year}{2007}).

\bibitem[{\citenamefont{Katsnelson}(2011)}]{Katsnelson2011}
\bibinfo{author}{\bibfnamefont{M.~I.} \bibnamefont{Katsnelson}},
  \bibinfo{journal}{Phys. Rev. B} \textbf{\bibinfo{volume}{84}},
  \bibinfo{pages}{041407} (\bibinfo{year}{2011}).

\bibitem[{\citenamefont{Hwang et~al.}(2011)\citenamefont{Hwang, Sensarma, and
  Das~Sarma}}]{Hwang2011}
\bibinfo{author}{\bibfnamefont{E.~H.} \bibnamefont{Hwang}},
  \bibinfo{author}{\bibfnamefont{R.}~\bibnamefont{Sensarma}}, \bibnamefont{and}
  \bibinfo{author}{\bibfnamefont{S.}~\bibnamefont{Das~Sarma}},
  \bibinfo{journal}{Phys. Rev. B} \textbf{\bibinfo{volume}{84}},
  \bibinfo{pages}{245441} (\bibinfo{year}{2011}).

\bibitem[{\citenamefont{Carrega et~al.}(2012)\citenamefont{Carrega,
  Tudorovskiy, Principi, Katsnelson, and Polini}}]{M.Carrega2012}
\bibinfo{author}{\bibfnamefont{M.}~\bibnamefont{Carrega}},
  \bibinfo{author}{\bibfnamefont{T.}~\bibnamefont{Tudorovskiy}},
  \bibinfo{author}{\bibfnamefont{A.}~\bibnamefont{Principi}},
  \bibinfo{author}{\bibfnamefont{M.~I.} \bibnamefont{Katsnelson}},
  \bibnamefont{and} \bibinfo{author}{\bibfnamefont{M.}~\bibnamefont{Polini}},
  \bibinfo{journal}{New Journal of Physics} \textbf{\bibinfo{volume}{14}},
  \bibinfo{pages}{063033} (\bibinfo{year}{2012}).

\bibitem[{\citenamefont{Narozhny et~al.}(2012)\citenamefont{Narozhny, Titov,
  Gornyi, and Ostrovsky}}]{Narozhny_drag_2012}
\bibinfo{author}{\bibfnamefont{B.~N.} \bibnamefont{Narozhny}},
  \bibinfo{author}{\bibfnamefont{M.}~\bibnamefont{Titov}},
  \bibinfo{author}{\bibfnamefont{I.~V.} \bibnamefont{Gornyi}},
  \bibnamefont{and} \bibinfo{author}{\bibfnamefont{P.~M.}
  \bibnamefont{Ostrovsky}}, \bibinfo{journal}{Phys. Rev. B}
  \textbf{\bibinfo{volume}{85}}, \bibinfo{pages}{195421}
  (\bibinfo{year}{2012}).

\bibitem[{\citenamefont{Kim and Tutuc}(2012)}]{Kim2012}
\bibinfo{author}{\bibfnamefont{S.}~\bibnamefont{Kim}} \bibnamefont{and}
  \bibinfo{author}{\bibfnamefont{E.}~\bibnamefont{Tutuc}},
  \bibinfo{journal}{Solid State Communications} \textbf{\bibinfo{volume}{152}},
  \bibinfo{pages}{1283} (\bibinfo{year}{2012}).

\bibitem[{\citenamefont{Gorbachev et~al.}(2012)\citenamefont{Gorbachev, Geim,
  Katsnelson, Novoselov, Tudorovskiy, Grigorieva, MacDonald, Morozov, Watanabe,
  Taniguchi et~al.}}]{Gorbachevi_nature_2012}
\bibinfo{author}{\bibfnamefont{R.~V.} \bibnamefont{Gorbachev}},
  \bibinfo{author}{\bibfnamefont{A.~K.} \bibnamefont{Geim}},
  \bibinfo{author}{\bibfnamefont{M.~I.} \bibnamefont{Katsnelson}},
  \bibinfo{author}{\bibfnamefont{K.~S.} \bibnamefont{Novoselov}},
  \bibinfo{author}{\bibfnamefont{T.}~\bibnamefont{Tudorovskiy}},
  \bibinfo{author}{\bibfnamefont{I.~V.} \bibnamefont{Grigorieva}},
  \bibinfo{author}{\bibfnamefont{A.~H.} \bibnamefont{MacDonald}},
  \bibinfo{author}{\bibfnamefont{S.~V.} \bibnamefont{Morozov}},
  \bibinfo{author}{\bibfnamefont{K.}~\bibnamefont{Watanabe}},
  \bibinfo{author}{\bibfnamefont{T.}~\bibnamefont{Taniguchi}},
  \bibnamefont{et~al.}, \bibinfo{journal}{Nat. Phys.}
  \textbf{\bibinfo{volume}{8}}, \bibinfo{pages}{896} (\bibinfo{year}{2012}).

\bibitem[{\citenamefont{Sch{\"u}tt et~al.}(2013)\citenamefont{Sch{\"u}tt,
  Ostrovsky, Titov, Gornyi, Narozhny, and Mirlin}}]{Sch_drag_prl_2013}
\bibinfo{author}{\bibfnamefont{M.}~\bibnamefont{Sch{\"u}tt}},
  \bibinfo{author}{\bibfnamefont{P.~M.} \bibnamefont{Ostrovsky}},
  \bibinfo{author}{\bibfnamefont{M.}~\bibnamefont{Titov}},
  \bibinfo{author}{\bibfnamefont{I.~V.} \bibnamefont{Gornyi}},
  \bibinfo{author}{\bibfnamefont{B.~N.} \bibnamefont{Narozhny}},
  \bibnamefont{and} \bibinfo{author}{\bibfnamefont{A.~D.}
  \bibnamefont{Mirlin}}, \bibinfo{journal}{Phys. Rev. Lett.}
  \textbf{\bibinfo{volume}{110}}, \bibinfo{pages}{026601}
  (\bibinfo{year}{2013}).

\bibitem[{\citenamefont{Gamucci et~al.}(2014)\citenamefont{Gamucci, Spirito,
  Carrega, Karmakar, Lombardo, Bruna, Pfeiffer, West, Ferrari, Polini
  et~al.}}]{Gamucci_nature_2014}
\bibinfo{author}{\bibfnamefont{A.}~\bibnamefont{Gamucci}},
  \bibinfo{author}{\bibfnamefont{D.}~\bibnamefont{Spirito}},
  \bibinfo{author}{\bibfnamefont{M.}~\bibnamefont{Carrega}},
  \bibinfo{author}{\bibfnamefont{B.}~\bibnamefont{Karmakar}},
  \bibinfo{author}{\bibfnamefont{A.}~\bibnamefont{Lombardo}},
  \bibinfo{author}{\bibfnamefont{M.}~\bibnamefont{Bruna}},
  \bibinfo{author}{\bibfnamefont{L.~N.} \bibnamefont{Pfeiffer}},
  \bibinfo{author}{\bibfnamefont{K.~W.} \bibnamefont{West}},
  \bibinfo{author}{\bibfnamefont{A.~C.} \bibnamefont{Ferrari}},
  \bibinfo{author}{\bibfnamefont{M.}~\bibnamefont{Polini}},
  \bibnamefont{et~al.}, \bibinfo{journal}{Nat. Commun.}
  \textbf{\bibinfo{volume}{5}} (\bibinfo{year}{2014}).

\bibitem[{\citenamefont{Titov et~al.}(2013)\citenamefont{Titov, Gorbachev,
  Narozhny, Tudorovskiy, Sch{\"u}tt, Ostrovsky, Gornyi, Mirlin, Katsnelson,
  Novoselov et~al.}}]{Hall_drag_Graphene}
\bibinfo{author}{\bibfnamefont{M.}~\bibnamefont{Titov}},
  \bibinfo{author}{\bibfnamefont{R.~V.} \bibnamefont{Gorbachev}},
  \bibinfo{author}{\bibfnamefont{B.~N.} \bibnamefont{Narozhny}},
  \bibinfo{author}{\bibfnamefont{T.}~\bibnamefont{Tudorovskiy}},
  \bibinfo{author}{\bibfnamefont{M.}~\bibnamefont{Sch{\"u}tt}},
  \bibinfo{author}{\bibfnamefont{P.~M.} \bibnamefont{Ostrovsky}},
  \bibinfo{author}{\bibfnamefont{I.~V.} \bibnamefont{Gornyi}},
  \bibinfo{author}{\bibfnamefont{A.~D.} \bibnamefont{Mirlin}},
  \bibinfo{author}{\bibfnamefont{M.~I.} \bibnamefont{Katsnelson}},
  \bibinfo{author}{\bibfnamefont{K.~S.} \bibnamefont{Novoselov}},
  \bibnamefont{et~al.}, \bibinfo{journal}{Phys. Rev. Lett.}
  \textbf{\bibinfo{volume}{111}}, \bibinfo{pages}{166601}
  (\bibinfo{year}{2013}).

\bibitem[{\citenamefont{Song and Levitov}(2013)}]{Song_Halldrag_2013}
\bibinfo{author}{\bibfnamefont{J.~C.~W.} \bibnamefont{Song}} \bibnamefont{and}
  \bibinfo{author}{\bibfnamefont{L.~S.} \bibnamefont{Levitov}},
  \bibinfo{journal}{Phys. Rev. Lett.} \textbf{\bibinfo{volume}{111}},
  \bibinfo{pages}{126601} (\bibinfo{year}{2013}).

\bibitem[{\citenamefont{Scharf and
  Matos-Abiague}(2012)}]{Scharf_Alex_drag_prb_2012}
\bibinfo{author}{\bibfnamefont{B.}~\bibnamefont{Scharf}} \bibnamefont{and}
  \bibinfo{author}{\bibfnamefont{A.}~\bibnamefont{Matos-Abiague}},
  \bibinfo{journal}{Phys. Rev. B} \textbf{\bibinfo{volume}{86}},
  \bibinfo{pages}{115425} (\bibinfo{year}{2012}).

\bibitem[{\citenamefont{Song et~al.}(2013)\citenamefont{Song, Abanin, and
  Levitov}}]{Drag_graphnen_nl}
\bibinfo{author}{\bibfnamefont{J.~C.~W.} \bibnamefont{Song}},
  \bibinfo{author}{\bibfnamefont{D.~A.} \bibnamefont{Abanin}},
  \bibnamefont{and} \bibinfo{author}{\bibfnamefont{L.~S.}
  \bibnamefont{Levitov}}, \bibinfo{journal}{Nano Letters}
  \textbf{\bibinfo{volume}{13}}, \bibinfo{pages}{36313637}
  (\bibinfo{year}{2013}).

\bibitem[{\citenamefont{Efimkin and Lozovik}(2013)}]{Drag_TI_2013_Efimkin}
\bibinfo{author}{\bibfnamefont{D.~K.} \bibnamefont{Efimkin}} \bibnamefont{and}
  \bibinfo{author}{\bibfnamefont{Y.~E.} \bibnamefont{Lozovik}},
  \bibinfo{journal}{Phys. Rev. B} \textbf{\bibinfo{volume}{88}},
  \bibinfo{pages}{235420} (\bibinfo{year}{2013}).

\bibitem[{\citenamefont{Liu et~al.}(2016{\natexlab{b}})\citenamefont{Liu, Liu,
  and Culcer}}]{Hong_drag_2015}
\bibinfo{author}{\bibfnamefont{H.}~\bibnamefont{Liu}},
  \bibinfo{author}{\bibfnamefont{W.~E.} \bibnamefont{Liu}}, \bibnamefont{and}
  \bibinfo{author}{\bibfnamefont{D.}~\bibnamefont{Culcer}},
  \bibinfo{journal}{Physica E: Low-dimensional Systems and Nanostructures}
  \textbf{\bibinfo{volume}{79}}, \bibinfo{pages}{7279}
  (\bibinfo{year}{2016}{\natexlab{b}}).

\bibitem[{\citenamefont{Murakami et~al.}(2004)\citenamefont{Murakami, Nagaosa,
  and Zhang}}]{Murakami_SHI_PRL04}
\bibinfo{author}{\bibfnamefont{S.}~\bibnamefont{Murakami}},
  \bibinfo{author}{\bibfnamefont{N.}~\bibnamefont{Nagaosa}}, \bibnamefont{and}
  \bibinfo{author}{\bibfnamefont{S.~C.} \bibnamefont{Zhang}},
  \bibinfo{journal}{Phys.\ Rev.\ Lett.} \textbf{\bibinfo{volume}{93}},
  \bibinfo{pages}{156804} (\bibinfo{year}{2004}).

\bibitem[{\citenamefont{K{\"o}nig et~al.}(2007)\citenamefont{K{\"o}nig,
  Wiedmann, Br{\"u}ne, Roth, Buhmann, Molenkamp, Qi, and
  Zhang}}]{QSH-HgTe_2007}
\bibinfo{author}{\bibfnamefont{M.}~\bibnamefont{K{\"o}nig}},
  \bibinfo{author}{\bibfnamefont{S.}~\bibnamefont{Wiedmann}},
  \bibinfo{author}{\bibfnamefont{C.}~\bibnamefont{Br{\"u}ne}},
  \bibinfo{author}{\bibfnamefont{A.}~\bibnamefont{Roth}},
  \bibinfo{author}{\bibfnamefont{H.}~\bibnamefont{Buhmann}},
  \bibinfo{author}{\bibfnamefont{L.~W.} \bibnamefont{Molenkamp}},
  \bibinfo{author}{\bibfnamefont{X.-L.} \bibnamefont{Qi}}, \bibnamefont{and}
  \bibinfo{author}{\bibfnamefont{S.-C.} \bibnamefont{Zhang}},
  \bibinfo{journal}{Science} \textbf{\bibinfo{volume}{318}},
  \bibinfo{pages}{766770} (\bibinfo{year}{2007}).

\bibitem[{\citenamefont{Chang et~al.}(2013)\citenamefont{Chang, Zhang, Feng,
  Shen, Zhang, Guo, Li, Ou, Wei, Wang et~al.}}]{Chang_QAHE_exper_Science2013}
\bibinfo{author}{\bibfnamefont{C.-Z.} \bibnamefont{Chang}},
  \bibinfo{author}{\bibfnamefont{J.}~\bibnamefont{Zhang}},
  \bibinfo{author}{\bibfnamefont{X.}~\bibnamefont{Feng}},
  \bibinfo{author}{\bibfnamefont{J.}~\bibnamefont{Shen}},
  \bibinfo{author}{\bibfnamefont{Z.}~\bibnamefont{Zhang}},
  \bibinfo{author}{\bibfnamefont{M.}~\bibnamefont{Guo}},
  \bibinfo{author}{\bibfnamefont{K.}~\bibnamefont{Li}},
  \bibinfo{author}{\bibfnamefont{Y.}~\bibnamefont{Ou}},
  \bibinfo{author}{\bibfnamefont{P.}~\bibnamefont{Wei}},
  \bibinfo{author}{\bibfnamefont{L.-L.} \bibnamefont{Wang}},
  \bibnamefont{et~al.}, \bibinfo{journal}{Science}
  \textbf{\bibinfo{volume}{340}}, \bibinfo{pages}{167} (\bibinfo{year}{2013}).

\bibitem[{\citenamefont{Chang et~al.}(2015)\citenamefont{Chang, Zhao, Kim,
  Zhang, Assaf, Heiman, Zhang, Liu, Chan, and Moodera}}]{Precise-QAHE-CZChang}
\bibinfo{author}{\bibfnamefont{C.-Z.} \bibnamefont{Chang}},
  \bibinfo{author}{\bibfnamefont{W.}~\bibnamefont{Zhao}},
  \bibinfo{author}{\bibfnamefont{D.~Y.} \bibnamefont{Kim}},
  \bibinfo{author}{\bibfnamefont{H.}~\bibnamefont{Zhang}},
  \bibinfo{author}{\bibfnamefont{B.~A.} \bibnamefont{Assaf}},
  \bibinfo{author}{\bibfnamefont{D.}~\bibnamefont{Heiman}},
  \bibinfo{author}{\bibfnamefont{S.-C.} \bibnamefont{Zhang}},
  \bibinfo{author}{\bibfnamefont{C.}~\bibnamefont{Liu}},
  \bibinfo{author}{\bibfnamefont{M.~H.~W.} \bibnamefont{Chan}},
  \bibnamefont{and} \bibinfo{author}{\bibfnamefont{J.~S.}
  \bibnamefont{Moodera}}, \bibinfo{journal}{Nat Mater}
  \textbf{\bibinfo{volume}{14}}, \bibinfo{pages}{473477}
  (\bibinfo{year}{2015}).

\bibitem[{\citenamefont{Mahoney et~al.}()\citenamefont{Mahoney, Colless, Pauka,
  Reilly, and et~al.}}]{D.Reilly_USYD}
\bibinfo{author}{\bibfnamefont{A.~C.} \bibnamefont{Mahoney}},
  \bibinfo{author}{\bibfnamefont{J.~I.} \bibnamefont{Colless}},
  \bibinfo{author}{\bibfnamefont{S.~J.} \bibnamefont{Pauka}},
  \bibinfo{author}{\bibfnamefont{D.~J.} \bibnamefont{Reilly}},
  \bibnamefont{and} \bibinfo{author}{\bibnamefont{et~al.}},
  \bibinfo{journal}{arXiv:1601.00634v}  (????).

\bibitem[{\citenamefont{Nagaosa et~al.}(2010)\citenamefont{Nagaosa, Sinova,
  Onoda, MacDonald, and Ong}}]{Nagaosa-AHE-2010}
\bibinfo{author}{\bibfnamefont{N.}~\bibnamefont{Nagaosa}},
  \bibinfo{author}{\bibfnamefont{J.}~\bibnamefont{Sinova}},
  \bibinfo{author}{\bibfnamefont{S.}~\bibnamefont{Onoda}},
  \bibinfo{author}{\bibfnamefont{A.~H.} \bibnamefont{MacDonald}},
  \bibnamefont{and} \bibinfo{author}{\bibfnamefont{N.~P.} \bibnamefont{Ong}},
  \bibinfo{journal}{Rev. Mod. Phys.} \textbf{\bibinfo{volume}{82}},
  \bibinfo{pages}{1539} (\bibinfo{year}{2010}).

\bibitem[{\citenamefont{Yao et~al.}(2008)\citenamefont{Yao, Xiao, and
  Niu}}]{Valley_prb_Niu}
\bibinfo{author}{\bibfnamefont{W.}~\bibnamefont{Yao}},
  \bibinfo{author}{\bibfnamefont{D.}~\bibnamefont{Xiao}}, \bibnamefont{and}
  \bibinfo{author}{\bibfnamefont{Q.}~\bibnamefont{Niu}},
  \bibinfo{journal}{Phys. Rev. B} \textbf{\bibinfo{volume}{77}},
  \bibinfo{pages}{235406} (\bibinfo{year}{2008}).

\bibitem[{\citenamefont{Yu et~al.}(2010)\citenamefont{Yu, Zhang, Zhang, Zhang,
  Dai, and Fang}}]{Yu_TI_QuantAHE_Science10}
\bibinfo{author}{\bibfnamefont{R.}~\bibnamefont{Yu}},
  \bibinfo{author}{\bibfnamefont{W.}~\bibnamefont{Zhang}},
  \bibinfo{author}{\bibfnamefont{H.~J.} \bibnamefont{Zhang}},
  \bibinfo{author}{\bibfnamefont{S.~C.} \bibnamefont{Zhang}},
  \bibinfo{author}{\bibfnamefont{X.}~\bibnamefont{Dai}}, \bibnamefont{and}
  \bibinfo{author}{\bibfnamefont{Z.}~\bibnamefont{Fang}},
  \bibinfo{journal}{Science} \textbf{\bibinfo{volume}{329}},
  \bibinfo{pages}{61} (\bibinfo{year}{2010}).

\bibitem[{\citenamefont{Ulstrup et~al.}(2014)\citenamefont{Ulstrup, Johannsen,
  Cilento, Miwa, Crepaldi, Zacchigna, Cacho, Chapman, Springate, Mammadov
  et~al.}}]{Bi-graphene_massive_prl}
\bibinfo{author}{\bibfnamefont{S.}~\bibnamefont{Ulstrup}},
  \bibinfo{author}{\bibfnamefont{J.~C.} \bibnamefont{Johannsen}},
  \bibinfo{author}{\bibfnamefont{F.}~\bibnamefont{Cilento}},
  \bibinfo{author}{\bibfnamefont{J.~A.} \bibnamefont{Miwa}},
  \bibinfo{author}{\bibfnamefont{A.}~\bibnamefont{Crepaldi}},
  \bibinfo{author}{\bibfnamefont{M.}~\bibnamefont{Zacchigna}},
  \bibinfo{author}{\bibfnamefont{C.}~\bibnamefont{Cacho}},
  \bibinfo{author}{\bibfnamefont{R.}~\bibnamefont{Chapman}},
  \bibinfo{author}{\bibfnamefont{E.}~\bibnamefont{Springate}},
  \bibinfo{author}{\bibfnamefont{S.}~\bibnamefont{Mammadov}},
  \bibnamefont{et~al.}, \bibinfo{journal}{Phys. Rev. Lett.}
  \textbf{\bibinfo{volume}{112}}, \bibinfo{pages}{257401}
  (\bibinfo{year}{2014}).

\bibitem[{\citenamefont{Lu et~al.}(2013)\citenamefont{Lu, Zhao, and
  Shen}}]{LuZhao_TITF_transport_PRL2013}
\bibinfo{author}{\bibfnamefont{H.-Z.} \bibnamefont{Lu}},
  \bibinfo{author}{\bibfnamefont{A.}~\bibnamefont{Zhao}}, \bibnamefont{and}
  \bibinfo{author}{\bibfnamefont{S.-Q.} \bibnamefont{Shen}},
  \bibinfo{journal}{Phys. Rev. Lett.} \textbf{\bibinfo{volume}{111}},
  \bibinfo{pages}{146802} (\bibinfo{year}{2013}).

\bibitem[{\citenamefont{Xiao et~al.}(2007)\citenamefont{Xiao, Yao, and
  Niu}}]{Valley-Hall_graphene_prl_2007}
\bibinfo{author}{\bibfnamefont{D.}~\bibnamefont{Xiao}},
  \bibinfo{author}{\bibfnamefont{W.}~\bibnamefont{Yao}}, \bibnamefont{and}
  \bibinfo{author}{\bibfnamefont{Q.}~\bibnamefont{Niu}},
  \bibinfo{journal}{Phys. Rev. Lett.} \textbf{\bibinfo{volume}{99}},
  \bibinfo{pages}{236809} (\bibinfo{year}{2007}).

\bibitem[{\citenamefont{Rojo}(1999)}]{1999_Hall_drag}
\bibinfo{author}{\bibfnamefont{A.~G.} \bibnamefont{Rojo}},
  \bibinfo{journal}{Journal of Physics: Condensed Matter}
  \textbf{\bibinfo{volume}{11}}, \bibinfo{pages}{R31} (\bibinfo{year}{1999}).

\bibitem[{\citenamefont{Tilahun et~al.}(2011)\citenamefont{Tilahun, Lee,
  Hankiewicz, and MacDonald}}]{Tilahun_TIF_QHS_PRL2011}
\bibinfo{author}{\bibfnamefont{D.}~\bibnamefont{Tilahun}},
  \bibinfo{author}{\bibfnamefont{B.}~\bibnamefont{Lee}},
  \bibinfo{author}{\bibfnamefont{E.~M.} \bibnamefont{Hankiewicz}},
  \bibnamefont{and} \bibinfo{author}{\bibfnamefont{A.~H.}
  \bibnamefont{MacDonald}}, \bibinfo{journal}{Phys. Rev. Lett.}
  \textbf{\bibinfo{volume}{107}}, \bibinfo{pages}{246401}
  (\bibinfo{year}{2011}).

\bibitem[{\citenamefont{Inoue et~al.}(2006)\citenamefont{Inoue, Kato, Ishikawa,
  Itoh, Bauer, and Molenkamp}}]{AHE_vertex_PRL_2006}
\bibinfo{author}{\bibfnamefont{J.-i.} \bibnamefont{Inoue}},
  \bibinfo{author}{\bibfnamefont{T.}~\bibnamefont{Kato}},
  \bibinfo{author}{\bibfnamefont{Y.}~\bibnamefont{Ishikawa}},
  \bibinfo{author}{\bibfnamefont{H.}~\bibnamefont{Itoh}},
  \bibinfo{author}{\bibfnamefont{G.~E.~W.} \bibnamefont{Bauer}},
  \bibnamefont{and} \bibinfo{author}{\bibfnamefont{L.~W.}
  \bibnamefont{Molenkamp}}, \bibinfo{journal}{Phys. Rev. Lett.}
  \textbf{\bibinfo{volume}{97}}, \bibinfo{pages}{046604}
  (\bibinfo{year}{2006}).

\bibitem[{\citenamefont{Hor et~al.}(2010)\citenamefont{Hor, Roushan,
  Beidenkopf, Seo, Qu, Checkelsky, Wray, Xia, Xu, Qian
  et~al.}}]{Hor_DopedTI_FM_PRB10}
\bibinfo{author}{\bibfnamefont{Y.~S.} \bibnamefont{Hor}},
  \bibinfo{author}{\bibfnamefont{P.}~\bibnamefont{Roushan}},
  \bibinfo{author}{\bibfnamefont{H.}~\bibnamefont{Beidenkopf}},
  \bibinfo{author}{\bibfnamefont{J.}~\bibnamefont{Seo}},
  \bibinfo{author}{\bibfnamefont{D.}~\bibnamefont{Qu}},
  \bibinfo{author}{\bibfnamefont{J.~G.} \bibnamefont{Checkelsky}},
  \bibinfo{author}{\bibfnamefont{L.~A.} \bibnamefont{Wray}},
  \bibinfo{author}{\bibfnamefont{Y.}~\bibnamefont{Xia}},
  \bibinfo{author}{\bibfnamefont{S.-Y.} \bibnamefont{Xu}},
  \bibinfo{author}{\bibfnamefont{D.}~\bibnamefont{Qian}}, \bibnamefont{et~al.},
  \bibinfo{journal}{Phys.\ Rev.\ B} \textbf{\bibinfo{volume}{81}},
  \bibinfo{pages}{195203} (\bibinfo{year}{2010}).

\bibitem[{\citenamefont{Cacho et~al.}(2015{\natexlab{b}})\citenamefont{Cacho,
  Crepaldi, Battiato, Braun, Cilento, Zacchigna, Richter, Heckmann, Springate,
  Liu et~al.}}]{Cacho_TI_tau}
\bibinfo{author}{\bibfnamefont{C.}~\bibnamefont{Cacho}},
  \bibinfo{author}{\bibfnamefont{A.}~\bibnamefont{Crepaldi}},
  \bibinfo{author}{\bibfnamefont{M.}~\bibnamefont{Battiato}},
  \bibinfo{author}{\bibfnamefont{J.}~\bibnamefont{Braun}},
  \bibinfo{author}{\bibfnamefont{F.}~\bibnamefont{Cilento}},
  \bibinfo{author}{\bibfnamefont{M.}~\bibnamefont{Zacchigna}},
  \bibinfo{author}{\bibfnamefont{M.~C.} \bibnamefont{Richter}},
  \bibinfo{author}{\bibfnamefont{O.}~\bibnamefont{Heckmann}},
  \bibinfo{author}{\bibfnamefont{E.}~\bibnamefont{Springate}},
  \bibinfo{author}{\bibfnamefont{Y.}~\bibnamefont{Liu}}, \bibnamefont{et~al.},
  \bibinfo{journal}{Phys. Rev. Lett.} \textbf{\bibinfo{volume}{114}},
  \bibinfo{pages}{097401} (\bibinfo{year}{2015}{\natexlab{b}}).

\bibitem[{\citenamefont{Peng et~al.}(2016)\citenamefont{Peng, Yang, Singh,
  Savrasov, and Yu}}]{Peng_natcom_2016_tau}
\bibinfo{author}{\bibfnamefont{X.}~\bibnamefont{Peng}},
  \bibinfo{author}{\bibfnamefont{Y.}~\bibnamefont{Yang}},
  \bibinfo{author}{\bibfnamefont{R.~R.~P.} \bibnamefont{Singh}},
  \bibinfo{author}{\bibfnamefont{S.~Y.} \bibnamefont{Savrasov}},
  \bibnamefont{and} \bibinfo{author}{\bibfnamefont{D.}~\bibnamefont{Yu}},
  \bibinfo{journal}{Nat Commun} \textbf{\bibinfo{volume}{7}},
  \bibinfo{pages}{10878} (\bibinfo{year}{2016}).

\bibitem[{\citenamefont{Vasko and Raichev}(2005)}]{Vasko}
\bibinfo{author}{\bibfnamefont{F.~T.} \bibnamefont{Vasko}} \bibnamefont{and}
  \bibinfo{author}{\bibfnamefont{O.~E.} \bibnamefont{Raichev}},
  \emph{\bibinfo{title}{Quantum Kinetic Theory and Applications}}
  (\bibinfo{publisher}{Springer}, \bibinfo{address}{New York},
  \bibinfo{year}{2005}).

\bibitem[{\citenamefont{Culcer}(2011)}]{Culcer_TI_Int_PRB11}
\bibinfo{author}{\bibfnamefont{D.}~\bibnamefont{Culcer}},
  \bibinfo{journal}{Phys. Rev. B} \textbf{\bibinfo{volume}{84}},
  \bibinfo{pages}{235411} (\bibinfo{year}{2011}).

\bibitem[{\citenamefont{Morimoto et~al.}(2015)\citenamefont{Morimoto, Furusaki,
  and Nagaosa}}]{TME_TI_PRB_Nagaosa}
\bibinfo{author}{\bibfnamefont{T.}~\bibnamefont{Morimoto}},
  \bibinfo{author}{\bibfnamefont{A.}~\bibnamefont{Furusaki}}, \bibnamefont{and}
  \bibinfo{author}{\bibfnamefont{N.}~\bibnamefont{Nagaosa}},
  \bibinfo{journal}{Phys. Rev. B} \textbf{\bibinfo{volume}{92}},
  \bibinfo{pages}{085113} (\bibinfo{year}{2015}).

\bibitem[{\citenamefont{Wehling et~al.}(2014)\citenamefont{Wehling,
  Black-Schaffer, and Balatsky}}]{DM_Balatsky_2014}
\bibinfo{author}{\bibfnamefont{T.}~\bibnamefont{Wehling}},
  \bibinfo{author}{\bibfnamefont{A.}~\bibnamefont{Black-Schaffer}},
  \bibnamefont{and} \bibinfo{author}{\bibfnamefont{A.}~\bibnamefont{Balatsky}},
  \bibinfo{journal}{Advances in Physics} \textbf{\bibinfo{volume}{63}},
  \bibinfo{pages}{1} (\bibinfo{year}{2014}).

\bibitem[{\citenamefont{Culcer et~al.}(2003)\citenamefont{Culcer, MacDonald,
  and Niu}}]{Culcer_AHE_PRB03}
\bibinfo{author}{\bibfnamefont{D.}~\bibnamefont{Culcer}},
  \bibinfo{author}{\bibfnamefont{A.~H.} \bibnamefont{MacDonald}},
  \bibnamefont{and} \bibinfo{author}{\bibfnamefont{Q.}~\bibnamefont{Niu}},
  \bibinfo{journal}{Phys.\ Rev.\ B} \textbf{\bibinfo{volume}{68}},
  \bibinfo{pages}{045327} (\bibinfo{year}{2003}).

\end{thebibliography}
\end{document}